\documentclass[11pt]{article}
\usepackage{amsmath,amsfonts,slashed}
\usepackage{tikz}
\usepackage{hyperref}
\usetikzlibrary{decorations.pathmorphing}
\newcommand{\eeB}{B^-}

\newcommand{\eeb}{b^-}

\newcommand{\zs}{{\declareslashed{}{\text{\textbackslash}}{0.04}{0}{z}\slashed{z}}}
\newcommand{\as}{{\declareslashed{}{\text{\textbackslash}}{0.04}{0}{a}\slashed{a}}}
\newcommand{\ts}{{\declareslashed{}{\text{\textbackslash}}{0.04}{0}{t}\slashed{t}}}
\newcommand{\Ts}{{\declareslashed{}{\text{\textbackslash}}{0.04}{0}{T}\slashed{T}}}
\newcommand{\mus}{{\declareslashed{}{\text{\textbackslash}}{0.04}{0}{\mu}\slashed{\mu}}}
\newcommand{\nus}{{\declareslashed{}{\text{\textbackslash}}{0.04}{0}{\nu}\slashed{\nu}}}
\newcommand{\ns}{{\declareslashed{}{\text{\textbackslash}}{0.04}{0}{n}\slashed{n}}}
\newcommand{\Ds}{{\declareslashed{}{\text{\textbackslash}}{0.04}{0}{D}\slashed{D}}}

\title{\bf\LARGE Singular Monopoles from Cheshire Bows}  

\author{\Large
Chris D. A. Blair\\
\\
\it School of Mathematics,\\
\it Trinity College, Dublin, Ireland\\
\tt cblair@maths.tcd.ie 
\and
\Large
Sergey A. Cherkis\thanks{On leave from {\it School of Mathematics and Hamilton Mathematics Institute,  Trinity College, Dublin, Ireland}}\\
\\
\it Department of Mathematics,\\
\it Stanford University, CA 94305\\
and\\
\it 
Department of Physics,\\ 
\it University of California,\\
\it Berkeley, CA 94720\\
\tt cherkis@maths.tcd.ie
}

\begin{document}
\begin{titlepage}

\renewcommand{\thepage}{ }
\date{}

\maketitle
\abstract{Singular monopoles are nonabelian monopoles with prescribed Dirac-type singularities.  All of them are delivered by the Nahm's construction.  In practice, however, the effectiveness of the latter is limited to the cases of one or two singularities.  We present an alternative construction of singular monopoles formulated in terms of Cheshire bows.  To illustrate the advantages of our bow construction we obtain an explicit expression for one $U(2)$ gauge group monopole with any given number of singularities of Dirac type.}

\vspace{-6in}

\parbox{\linewidth}
{\small\hfill \shortstack{TCDMATH 10-08\\ \hfill HMI 10-05}}

\end{titlepage}

\tableofcontents

\section{Introduction}
We formulate a new construction of singular monopoles and illustrate its every step by explicitly computing one monopole with $k$ Dirac-type singularities as an example.  Until now the conventional techniques were limited to $k=1$ and $k=2$ cases.   Our construction is equally effective for any number of singularities.   The elements of our construction are conveniently organized in terms of bows, which are generalizations of quivers, introduced in \cite{Cherkis:2010bn,Cherkis:2009jm,Cherkis:2008ip}.
Originally bows were introduced in order to find Yang-Mills instantons on curved backgrounds of asymptotically locally flat gravitational instantons.  As we argue here, by restricting attention in this bow construction to what we call Cheshire bow representations one obtains an alternative way of finding all singular monopoles.

\subsection{The Use of Singular Monopoles}
Singular monopoles play an important role in a number of physical problems and have diverse mathematical applications.  These classical Yang-Mills-Higgs configurations are directly related to
\begin{itemize}
\renewcommand{\labelitemi}{$\cdot$}
\item the vacua and the low energy behavior of supersymmetric gauge  theories in three dimensions, 
\item the electric-magnetic duality of maximally supersymmetric Yang-Mills in four space-time dimensions, 
\item Yang-Mills instantons on curved backgrounds,
\item string theory brane configurations, and
\item gravitational instantons.    
\end{itemize}

As first suggested in \cite{Chalmers:1996xh} and explored in e.g. \cite{Seiberg:1996nz,Hanany:1996ie,Cherkis:1997aa}, the moduli spaces of vacua of the quantum three-dimensional ${\cal N}=4$ supersymmetric gauge theories are given by the moduli spaces of singular monopoles.  In particular the quantum moduli space of vacua of the ${\cal N}=4$ $U(n)$ super-Yang-Mills theory with $k$ matter hypermultiplets in the fundamental representation is the classical moduli space of $U(2)$ monopoles of nonabelian charge $n$ with $k$ minimal singularities.   
In the exploration \cite{Borokhov:2003yu,Kapustin:2005py} of the Montonen-Olive duality \cite{Montonen:1977sn}, or more exactly its supersymmetric version \cite{Witten:1978mh},  the Goddard-Nuyts-Olive (GNO) singularities \cite{Goddard:1976qe} of the type we study here represent 't Hooft operators that are dual to the Wilson operators.  In fact it is the study of the monopole singularities in \cite{Goddard:1976qe} that prompted the discovery of the electric-magnetic duality \cite{Montonen:1977sn}.  On the other hand, it was demonstrated in \cite{Kapustin:2006pk} that one of the consequences of the electric-magnetic duality of the maximally supersymmetric Yang-Mills theory is the geometric Langlands correspondence.  As a result, singular monopoles are significant in the study of the geometric Langlands duality; in particular, in \cite{Kapustin:2006pk} the moduli spaces of singular monopoles were identified with the spaces of Hecke transformations.  Such a close relationship was also observed in \cite{Baptista:2009zb}.  

There is a very close connection between monopoles and instantons.  For example an instanton on a space with a periodic direction, called a caloron, can be thought of as a nonlinear superposition of monopoles and antimonopoles \cite{Kraan:1998pm,Bruckmann:2003yq}.  In a different view \cite{Garland:1988bv,Garland:1989cd} a caloron with a gauge group $G$ can be thought as a monopole with the loop group of $G$ as its structure group.  One can envisage an extension of these results to instantons on a multi-Taub-NUT space ($TN_k$) with $k$ Taub-NUT centers. We conjecture that the corresponding generalization of the former statement is that an instanton on $TN_k$ is a nonlinear superposition of  singular monopoles and antimonopoles.  And the analogue of the latter statement is that an instanton on $TN_k$ with a gauge group $G$ is a singular monopole with the loop group of $G$ as its structure group.

Singular monopoles describe Chalmers-Hanany-Witten brane configurations of the type IIB string theory \cite{Chalmers:1996xh,Hanany:1996ie} and are very useful in exploring their various properties.  In \cite{Cherkis:1998hi} they were instrumental in obtaining the twistor spaces of Gravitational Instantons, metrics on which were found in \cite{Cherkis:2003wk}.

The twistor theory and the moduli spaces of  singular monopoles were first studied in \cite{Kronheimer}.  In particular the moduli space of one $U(2)$ monopole with $k$ minimal singularities, which is the configuration we explicitly obtain here, is the $k$-centered multi-Taub-NUT space \cite{Kronheimer}.  The centered moduli space of two $U(2)$ monopoles with $k$ singularities is the $D_k$ ALF space \cite{Cherkis:1998hi,Cherkis:2003wk}.

These are some of the uses of singular monopoles.  Now we turn describing the singular monopole configurations and their construction. 

\subsection{Singular Monopole Constructions}
By a BPS monopole \cite{Bogomolny:1975de,Prasad:1975kr}
\footnote{Normally one requires a monopole to have finite energy $\int_{\mathbb{R}^3} {\rm tr} (F\wedge*F+D\Phi\wedge*D\Phi).$ For singular monopoles, however, this condition is relaxed. Instead one excises small balls $B_j$ centered around the points $\nu_j$ and requires the energy outside $\int_{\mathbb{R}^3\setminus\, \cup_jB_j} {\rm tr} (F\wedge*F+D\Phi\wedge*D\Phi)$ to be finite, while the singularity inside each ball $B_j$ is prescribed.}  
we understand a pair $(A, \Phi)$ of a hermitian connection $A$ and a hermitian Higgs field $\Phi$ satisfying the Bogomolny equation
\begin{equation}
F_{ab}+\sum_{c=1}^3\epsilon_{abc}[D_c,\Phi]=0,
\end{equation}
where $F$ is the curvature of $A.$  Using differential forms this equation is written as $F+*D\Phi=0,$ where $*$ is the Hodge start operator.  A {\em singular monopole} with singularities at points $\vec{\nu}_j\in\mathbb{R}^3,\ j=1,\ldots,k$ is a BPS monopole with $A$ and $\Phi$ regular everywhere except at points $\vec{\nu}_j,$ where locally they are required to have the prescribed behavior
\begin{align}
\Phi\big(\vec{t}\:\big)&=\frac{\left(1+\ns\right)}{4\big|\vec{t}-\vec{\nu}_j\big|}+O\left(\big|\vec{t}-\vec{\nu}_j\big|^0\right),&
A\big(\vec{t}\:\big)&=\frac{1+\ns}{2}\omega_j+O\left(\big|\vec{t}-\vec{\nu}_j\big|^0\right).
\end{align}
Here $\vec{n}=(n_1,n_2,n_3)$ is a unit vector and we are using the notation $\ns = n_1 \sigma_1 + n_2 \sigma_2 + n_3 \sigma_3$ with $\sigma_1, \sigma_2,\sigma_3$ the Pauli matrices. This is exactly the Dirac monopole at each $\vec{\nu}_j$ embedded into the gauge group $U(2)$ with, for example, $\omega_j=-\frac{(\vec{T}\times\vec{t}_j)\cdot d\vec{t}}{2t_j(Tt_j-\vec{T}\cdot\vec{t}_j)}$ for some choice of $\vec{T}.$

The technique for constructing a general regular monopole was discovered by Nahm \cite{Nahm:1979yw, NahmCalorons}.  For a $U(2)$ monopole with $k$ singularities this technique was used in \cite{Cherkis:1998hi,Cherkis:1998xca}  to study the metric on their moduli space.  The starting point of the Nahm's construction of singular monopoles is a solution of the Nahm equations either on a real line or on a semi-infinite interval.  While being very efficient in the study of the moduli spaces, it would be difficult to apply this construction if one is to find the monopole configurations themselves for arbitrary number of singularities. For the case of one or two singularities this construction is tractable and was employed in \cite{Cherkis:2007qa, Cherkis:2007jm} producing explicit solutions. Unfortunately, for a more general case, the difficulty is that the Nahm data, which is the starting point of the construction, contains a rank $k$ solution of the Nahm equations on a semi-infinite interval.  For $k>2$ such solutions are difficult to construct and to work with. 

In order to circumvent this difficulty, we shall employ the novel technique of bow diagrams introduced in \cite{Cherkis:2008ip} and developed in \cite{Cherkis:2010bn,Cherkis:2009jm}.  Bow diagrams were introduced in order to construct all instantons, i.e. solutions of the Yan-Mills self-duality equation, on the multi-Taub-NUT space $TN_k.$ All such instantons of  given charges are given by a bow representation of the $A_{k-1}$ bow, also called $TN_k$ bow, such as in Figure~\ref{TheBow}.  A representation is determined by a collection of points on a bow and the ranks of bundles over the intervals between these points. The positions of these points correspond to the eigenvalues of the Polyakov loop at infinity of $TN_k,$ while the bundle ranks determine the charges.  

What does the bow construction for instantons has to do with the singular monopole problem we are considering here? In \cite{Kronheimer} Kronheimer observed that any self-dual connection on a $k$-centered multi-Taub-NUT space that is invariant under the triholomorphic isometry  of the multi-Taub-NUT space is equivalent to a solution of the Bogomolny equation $F=-*D\Phi$ on $\mathbb{R}^3$, with $k$ singularities corresponding to the Taub-NUT center locations.  Thus our problem of singular monopoles with $k$ singularities is equivalent to the problem of $\theta$-independent instantons on $TN_k.$ In terms of the bow representation the condition that guarantees the invariance of the resulting solution under the isometry is that one of the ranks determining the bow representation is zero. We call such a representation a {\em Cheshire representation}.  This is exactly what one needs to find the singular monopole solutions we seek.  As a matter of fact this representation provides a general construction for singular monopoles of any charge.  

In the following sections we present the $A_{k-1}$ bow and explain its relation to the multi-Taub-NUT space and abelian instantons on it. In section \ref{Sec:Cheshire} we identify the relevant Cheshire representations of the bow and its data, and outline the transform of  \cite{Cherkis:2009jm} which in this case produces singular monopole solutions. We then apply this transform to obtain one generic $U(2)$ monopole solution with $k$  minimal singularities positioned at $\vec{\nu}_j,\ j=1,2,\ldots,k.$  
\begin{figure}[ht]
\begin{center}
\includegraphics[width=0.6\textwidth]{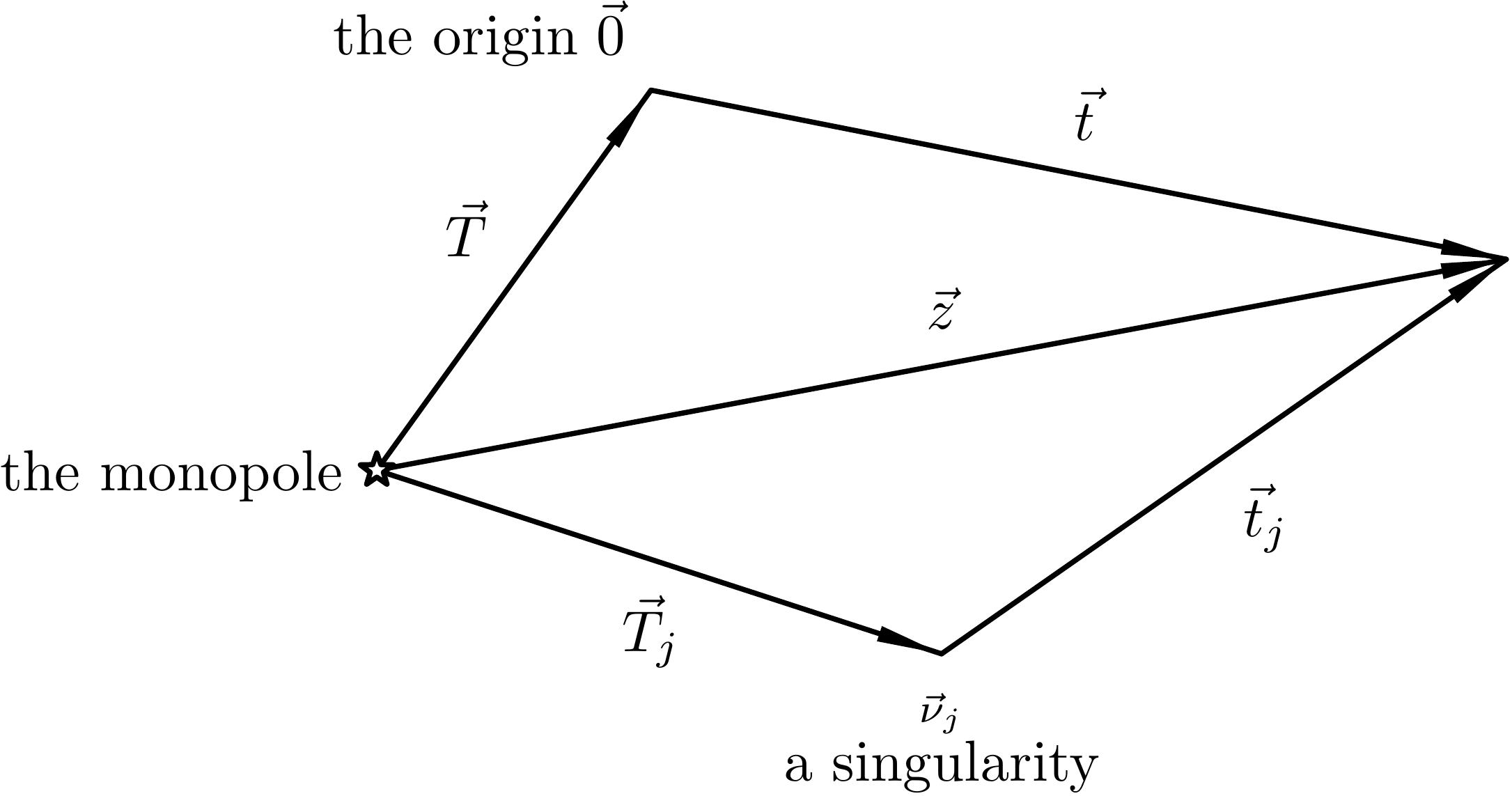}
\caption{The relative positions of the observation point $\vec{t},$ the monopole $-\vec{T},$ and one of the singularities $\vec{\nu}_j.$  The distances are $t_j=|\vec{t}-\vec{\nu}_j|,  T_j=|\vec{T}+\vec{\nu}_j|,$ and $z=|\vec{t}+\vec{T}|.$ }
\label{Constellation}
\end{center}
\end{figure}We find its Higgs field and connection to have a relatively simple form:
\begin{multline}
\Phi\Big(\vec{t}\,\Big) =  \left( \left[ \lambda + \sum_{j=1}^k \frac{1}{4t_j}  \right] \coth 2(\lambda + \alpha) z -\frac{1}{2z} \right) \frac{\zs}{z}\\
 + \frac{z}{ \sinh 2 (\lambda + \alpha ) z }\sum_{j=1}^k\frac{\Ts_{j\,\perp}}{2t_j \big((T_j+t_j)^2-z^2\big)}
 +\sum_{j=1}^k \frac{1}{4t_j}, 
\end{multline}
\begin{multline}
A\Big(\vec{t}\, \Big)  =\left(\frac{1}{2z}-\frac{1}{\sinh 2(\lambda+\alpha)z} \left[ \lambda + \sum_{j=1}^k  \frac{T_j+t_j}{2\big((T_j+t_j)^2-z^2\big)}\right]\right) \frac{i[\zs, d\ts]}{2z}\\
     + \frac{z}{\sinh 2 (\lambda + \alpha) z} \sum_{j=1}^k \frac{i[\ts_j, d\ts]_\perp}{4t_j\big((T_j+t_j)^2-z^2\big) }\\
     -\left(1+ \frac{\zs}{z} \coth 2(\lambda +\alpha)z \right)\sum_{j=1}^k \frac{(\vec{T}_j\times\vec{t}_j)\cdot d\vec{t}}{2 t_j((T_j+t_j)^2-z^2)},
\end{multline}
where the function $\alpha$ is given by 
\begin{equation}
\exp(4\alpha z)=\prod_j\frac{T_j+t_j+z}{T_j+t_j-z}.
\end{equation}
The eigenvalues of the Higgs field at infinity are $\pm\lambda$ and $-\vec{T}$ determines the position of the nonabelian monopole, as in Figure~\ref{Constellation}. 

We would like to emphasize that the Cheshire bow construction we formulate here delivers all singular monopoles.  We focus on one singular monopole as an illustrative example making every detail explicit.

\section{Cheshire Bow Construction}
The core idea of this work combines the observation of Kronheimer relating singular monopoles with instantons on multi-Taub-NUT space together with the bow construction of such instantons.  Let us begin by formulating the conventional Nahm transform for singular monopoles and highlighting the technical difficulties one faces in its practical application.  Then we proceed by presenting Kronheimer's relation and formulating our generalization of the Nahm transform.  This gives an alternative construction of singular monopoles.
\subsection{The Nahm Transform}\label{NahmTransform}
In order to construct a $U(2)$ monopole of nonabelian charge $m$ with $k$ singularities using the conventional Nahm transform one begins by finding the Nahm data $(T_1(s),T_2(s),T_3(s))$ consisting of three hermitian matrix valued functions of one variable $s$ that satisfy the Nahm equations
\begin{align}
\frac{d}{ds}T_1&=i [T_2, T_3],\\
\frac{d}{ds}T_2&=i [T_3, T_1],\\
\frac{d}{ds}T_3&=i [T_1, T_2].
\end{align}
If the asymptotic eigenvalues of the monopole Higgs field we are constructing are $\lambda_1$ and $\lambda_2$ with $\lambda_1<\lambda_2,$ then the Nahm data is of rank $m$ on the interval $[\lambda_1,\lambda_2]$ and rank $k$ on the semi-infinite interval $(\lambda_2,+\infty).$ For concreteness, let us presume that $k>m,$  then at $\lambda_2$ the matching  condition states that the smaller rank $T$ is a block in of the larger rank $T,$ so that, for $s>\lambda_2$
\begin{equation}
T_a(s)=\left(\begin{array}{cc}
\frac{\rho_a}{s-\lambda_2}+O(1) &O\Big((s-\lambda_2)^{\frac{k-m-1}{2}}\Big) \\
O\Big((s-\lambda_2)^{\frac{k-m-1}{2}}\Big) & T_a(\lambda_2)+O(s-\lambda_2)
\end{array}\right),
\end{equation}
where the residues $\rho_1, \rho_2,$ and $\rho_3$ satisfy $[\rho_a,\rho_b]=\sum_c\epsilon_{abc}i\rho_c,$ forming a $(k-m)$-dimensional irreducible representation of $su(2)$ generators.  The condition at $\lambda_1$ is that 
\begin{equation}
T_a(s)=\frac{\rho'_a}{s-\lambda_1},
\end{equation}
with $\rho'_a$ forming an $m$-dimensional irreducible  representation of the $su(2)$ generators.  If the positions of the monopole singularities are $\vec{\nu}_j,$ then the conditions one imposes on the eigenvalues of the  Nahm data at $s=\infty$ are 
\begin{equation}
\lim_{s\rightarrow+\infty}\text{EigVal}\ T_a(s)=\text{diag}(\nu^a_1, \nu^a_2, \ldots, \nu^a_k).
\end{equation}

Given any such solution $(T_1, T_2, T_3)$ Nahm constructs a family of Dirac (or Weyl) operators parameterized by $\vec{t}\in\mathbb{R}^3$:
$\Ds=-\frac{d}{ds}-\Ts-\ts,$
and a family of conjugate operators 
\begin{equation}\label{NahmDirac}
\Ds^\dagger=\frac{d}{ds}-\Ts-\ts.
\end{equation} 
These operators act on $L^2$ fundamental spinors over the interval  $(\lambda_1,+\infty).$  All such fundamental spinor-valued functions form a trivial bundle over the $\mathbb{R}^3$ parameterized by $\vec{t},$ and the kernel of $\Ds^\dagger$ is a subbundle of this trivial bundle.  For each value of $\vec{t}$ the kernel is two dimensional.  If $\psi_1(s,\vec{t}\:)$ and $\psi_2(s,\vec{t}\:)$ form an orthonormal basis of this kernel, then one forms the Higgs field $\Phi=(\Phi_{\alpha\beta})$ and the connection $A=(A_{\alpha\beta})$ with the components
\begin{align}
\Phi_{\alpha\beta}\big(\vec{t}\:\big)&=\int_{\lambda_1}^{+\infty}s\psi_\alpha^\dagger\psi_\beta \,ds,&
A^a_{\alpha\beta}\big(\vec{t}\:\big)&=i\int_{\lambda_1}^{+\infty}\psi_\alpha^\dagger\frac{\partial}{\partial t^a}\psi_\beta \,ds,
\end{align}
which together constitute a singular monopole.  This is the conventional Nahm transform \cite{Nahm:1979yw, NahmCalorons} as formulated in \cite{Cherkis:1998xca}.  For every gauge equivalence class of solutions of the Nahm equations with the boundary conditions specified above it produces a $U(2)$ singular monopole with minimal singularities at $\vec{\nu}_j$ and nonabelian charge $m.$

This transform was successfully applied to find singular monopoles with one \cite{Cherkis:2007jm} and two singularities \cite{Cherkis:2007qa}.  
As we already pointed out, it is substantially more difficult, though not impossible, to use for a larger number of singularities.  This is one of the reasons we proceed to introduce an alternative construction of singular monopoles, which we now outline.

\subsection{Kronheimer's Correspondence}
The multi-Taub-NUT space is a four-dimensional space with the metric
\begin{equation}\label{Eq:mTN}
ds^2=Vd\vec{t}\:^2+\frac{(d\theta+\omega)^2}{V},
\end{equation}
with $\theta$ of period $2\pi,$ $V=l+\sum_{j=1}^k\frac{1}{2|\vec{t}-\vec{\nu}_j|},$ and $d\omega=-*_3 dV.$ 
 A Yang-Mills connection $\hat{A}$ on this space can be written in the form 
 \begin{equation}
\hat{A}=A-\Phi\frac{d\theta+\omega}{V}.
\end{equation}
As observed in \cite{Kronheimer}, if this connection satisfies the self-duality equation on the multi-Taub-NUT space and if there is a gauge transformation that makes $A$ and $\Phi$ $\theta$-independent, then we can understand the fields $A$ and $\Phi$ as a connection and a Higgs field on $\mathbb{R}^3$ satisfying the Bogomolny equation
\begin{equation}
F_A+*[D_A,\Phi]=0.
\end{equation}
If before the gauge transformation the field $\hat{A}$ was smooth and had a finite action, then the resulting configuration $(A,\Phi)$ is a singular monopole with singularities at the positions of the Taub-NUT centers $\vec{\nu}_j.$  It is the action of this gauge transformation at the points $\vec{\nu}_j$ that determines the charges of the singularities \cite{Kronheimer}. 

With this in mind, instead of searching for singular monopoles we can try to solve an equivalent, though at first sight more complicated looking, problem  of finding instantons on the multi-Taub-NUT space that are $\theta$-independent.

\subsection{Bows and Instantons on multi-Taub-NUT}
\begin{figure}[htbp]
\begin{center}
\includegraphics[width=0.4\textwidth]{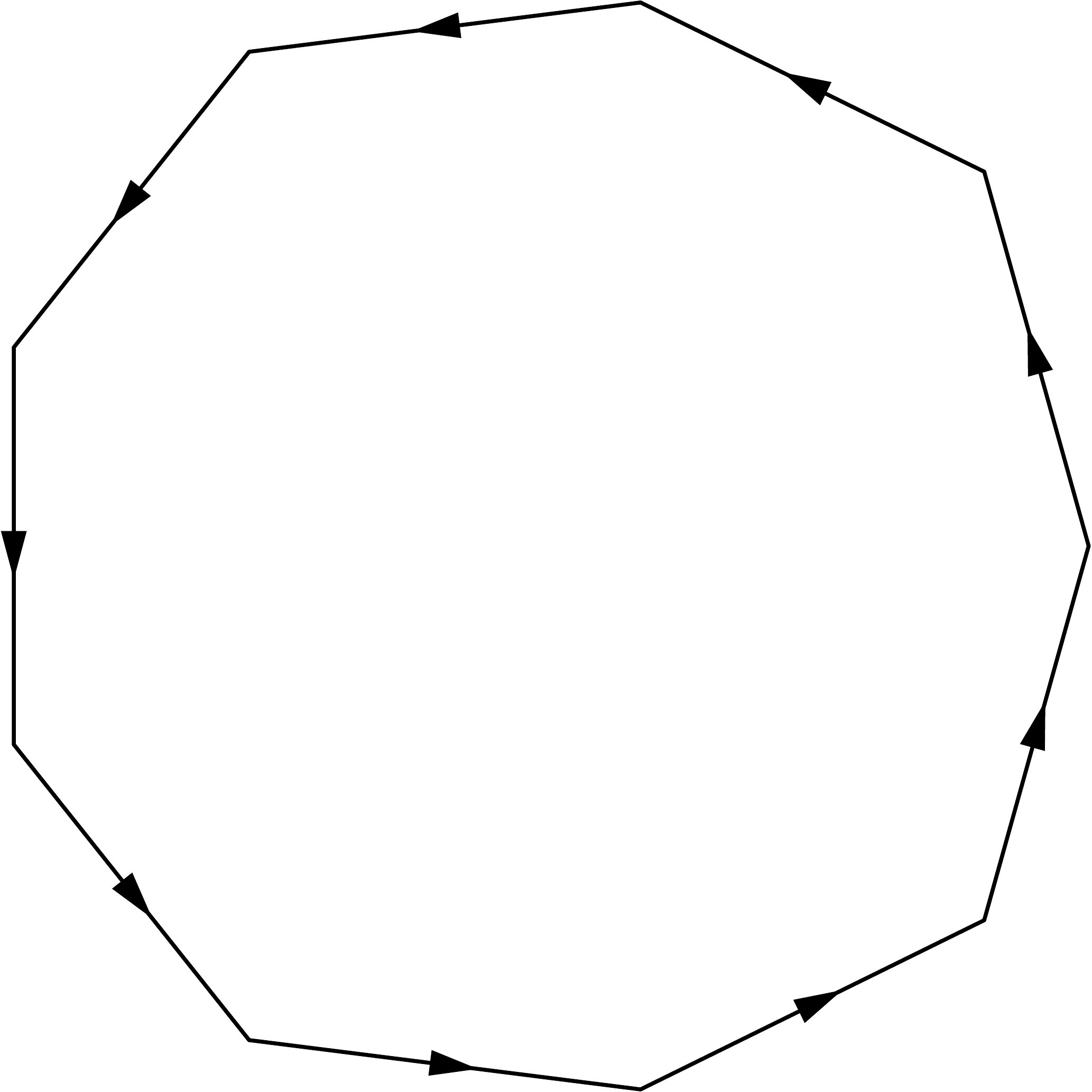}
\caption{An example of the affine $A_{k-1}$ quiver.  This is an $A_8$ affine quiver giving the $A_8$ ALE space and instantons on it.}
\label{Quiver}
\end{center}
\end{figure}

A multi-Taub-NUT space with $k$ Taub-NUT centers is a close cousin of the $A_{k-1}$ Asymptotically Locally Euclidean (ALE) space.  This space is given by the metric \eqref{Eq:mTN} with the parameter $l=0.$ The asymptotic form of its metric approaches the flat metric on $\mathbb{R}^4/\mathbb{Z}_k.$  The instantons on the $A_{k-1}$ ALE space, and on all ALE spaces, were constructed by Kronheimer and Nakajima \cite{KN}.  This construction is formulated in terms of quivers.  The relevant quiver is the affine $A_{k-1}$ quiver, such as the one in Figure~\ref{Quiver}.

The recent construction of instantons on multi-Taub-NUT spaces \cite{Cherkis:2010bn,Cherkis:2009jm,Cherkis:2008ip} generalizes the notion of quivers to the notion of bows.  If a quiver consists of points and oriented edges connecting them, a bow consists of intervals and oriented edges connecting them. We refer to \cite{Cherkis:2010bn} for the exact definitions.  An $A_{k-1}$ bow appears in Figure \ref{TheBow}.  It has various representations, each representation of a bow corresponding to a class of all instantons with given topological charges.  A representation of a bow is a collection of points $\lambda_\alpha$ belonging to its intervals and a collection of vector bundles over the subintervals into which these intervals are divided by the $\lambda$-points.  In particular some of these bundles can have rank zero, in which case their corresponding subintervals play no role and do not contribute to the final instanton connection.  If this is indeed the case and a representation has at least one of its bundles of rank zero we call it a {\em Cheshire representation}.

Now, among all of the bow representations it remains to single out those that produce self-dual connections that are $\theta$-independent.  How does the $\theta$ dependence arise?  

To implement this construction one needs two representations of the same $TN_k$ bow. We call them large and small representations.  A data of the large representation determines the instanton, while the data of the small representation parameterizes the multi-Taub-NUT space.
For a small representation on each of the bow intervals one considers the Nahm data consisting of the abelian $U(1)$ connection $t_0$ and three abelian Higgs fields $t_1, t_2, t_3.$  The three Higgs fields give rise to the three of the multi-Taub-NUT coordinates assembled into a vector $\vec{t},$ while the coordinate $\theta$ is the logarithm of the Polyakov loop $\int t_0(s)ds.$  Our construction is gauge invariant and therefore we can locally adjust the values of $t_0$, even gauging it away on some intervals completely.  The only objects that remains invariant under the gauge transformations are the Polyakov loop and $t_1, t_2,$ and $t_3.$  Given the large bow  representation data we form a family of operators similar to the $\Ds^\dagger$ operators of Eq.~\eqref{NahmDirac} that appeared in the conventional Nahm transform of Section~\ref{NahmTransform}.  These operators depend only on the values of $t_0$ on the subintervals where the rank of the large representation bundle is nonzero.  Therefore, if all ranks of the large representation are positive, then the resulting connection does depend on $t_0$ and therefore on $\theta.$  If one of the ranks is zero, however, then we can work in a gauge where $t_0$ is gauged away on all sub-intervals, except the one carrying the zero rank bundle. As a result the kernel of our operators will be independent of $\theta$ and so will be the resulting connection.

\section{The Multi-Taub-NUT Space}
\begin{figure}[htbp]
\begin{center}
\includegraphics[width=0.4\textwidth]{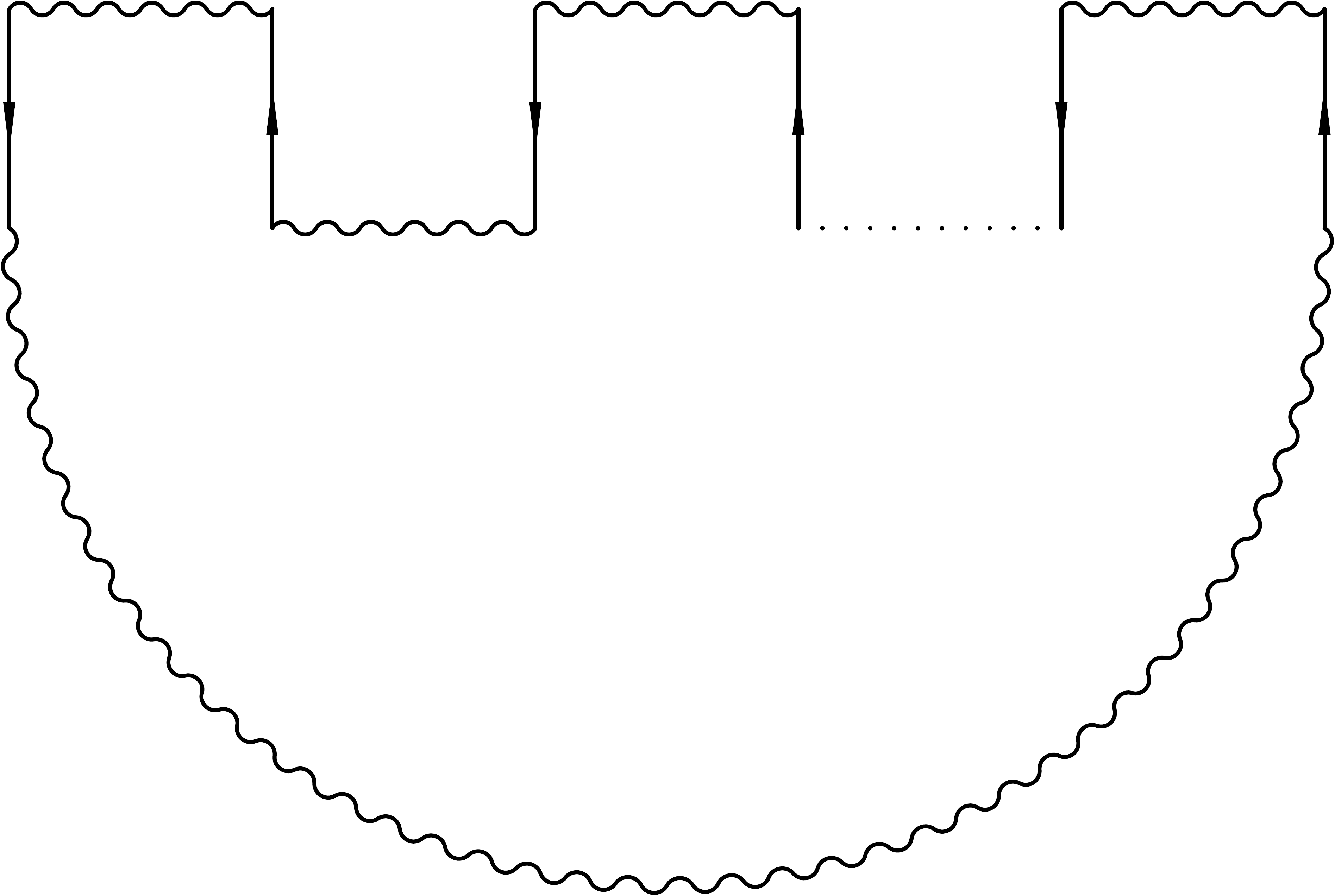}
\caption{$A_{k-1}$ Bow.  It has multi-Taub-NUT space with $k$ centers as the moduli space of its small representation.  Any other representation of this bow delivers self-dual connections on this multi-Taub-NUT space.}
\label{TheBow}
\end{center}
\end{figure}

A general definition of a bow, its representation, and its data can be found in \cite{Cherkis:2010bn}.  Here we focus on the $A_{k-1}$  bow, also called the  $TN_k$ bow, given in Figure \ref{TheBow}.  
It consists of $k$ intervals $I_j, j=1,\ldots,k$ denoted by the wavy lines and $k$ oriented  edges denoted by the arrows connecting the ends of the wavy lines.  We parameterize the intervals by the variable $s,$ and for concreteness denote the left end of $I_j$ by $p_j^L$ and the right end by $p_j^R$ so that $I_j=[p_j^L, p_j^R].$  In what follows we can understand the variable $s$ to be parameterizing a circle of circumference $l.$  This circle is divided into intervals $I_j,$ and even though in this picture any two neighboring intervals $I_{j-1}$ and $I_j$ appear to share an endpoint, we still treat the ends of any two intervals $p_{j-1}^R$ and $p_j^L$ as distinct points. One of the simplest representations of this bow has rank one bundles on each interval and no $\lambda$-points.  We call this the {\em small representation} and denote the associated data by small letters $t$ and $b.$  Let us begin by discussing this representation in detail and by finding its moduli space. 

Each interval $I_j$ has an associated line bundle $e_j\rightarrow I_j$ with connection $\frac{d}{ds}-i t_0(s)$ and three Higgs fields $t_1(s), t_2(s), t_3(s).$ 
Each edge, say the $j^{\rm th}$ edge, connects the intervals $j-1$ and $j$ as in Figure \ref{fig:edge}, with the tail $t(j)$ being the right end of the $(j-1)^{\rm st}$ interval, $p_j^L=h(j),$ and the head $h(j)$ being the left end of the $j^{\rm th}$ interval, $p_{j-1}^R=t(j).$ If $e_{t(j)}$ denotes the fiber of $e_{j-1}$ at the right end of the interval $I_{j-1}$ and $e_{h(j)}$ denotes the fiber of the bundle $e_{j}$ at the left end of the interval $I_{j},$ then we consider linear maps
\begin{align}
b_j^{LR}:\,\,& e_{t(j)}\rightarrow e_{h(j)}& &\text{and}&
b_j^{RL}:\,\,& e_{h(j)}\rightarrow e_{t(j)},
\end{align}
associated with the $j^\text{th}$ edge.

\begin{figure}[h]\centering
\begin{tikzpicture}
\begin{scope}[xshift=-1cm,thick]
\draw[ decorate,decoration={snake,amplitude=.4mm,segment length=2.5mm}] (-2,0)--(0,0);
\draw[ decorate,decoration={snake,amplitude=.4mm,segment length=2.5mm}] (0,2)--(2,2);
\draw[->] (0,0) -- (0,1);
\draw (0,1) -- (0,2);
\draw (-1,-0.25) node {\scriptsize $j-1$};
\draw (1,1.7) node {\scriptsize $j$};
\draw (0,2.25) node {\scriptsize $h(j)$};
\draw (2,2.25) node {\scriptsize $$};
\draw (-2,-0.25) node {\scriptsize $$};
\draw (0,-0.25) node {\scriptsize $t(j)$};
\draw (-0.85,1.35) node {\scriptsize edge $j$};
\end{scope}
\end{tikzpicture}
\caption{An edge}
\label{fig:edge}
\end{figure}
\begin{figure}[htbp]
\begin{center}
\includegraphics[width=0.5\textwidth]{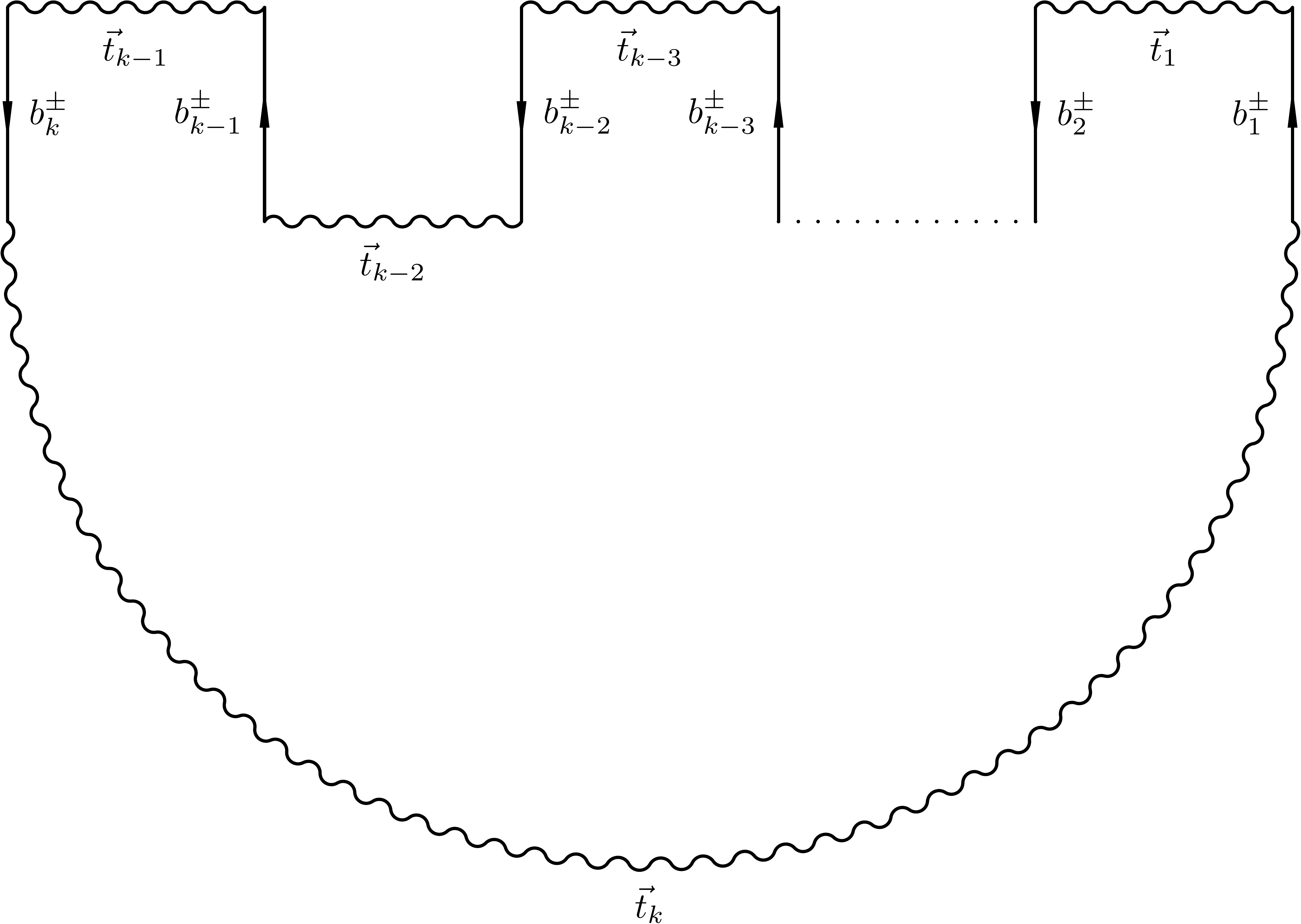}
\caption{Small Bow Representation: This bow has $k$ intervals.  Assigning a line bundle to each defines a representation with $k$-centered Taub-NUT as its moduli space.}
\label{SmallRep}
\end{center}
\end{figure}

These are assembled into $b_j^+$ and $\eeb_j$ as
\begin{align}
b_j^+&=\left(\begin{array}{c}  \overline{b_j^{RL}}\\ -b_j^{LR}  \end{array}\right)& &\text{and}&
\eeb_j &=\left(\begin{array}{c}  \overline{b_j^{LR}}\\ b_j^{RL}  \end{array}\right).
\end{align}
Figure~\ref{SmallRep} assembles all this data into a decorated bow. 
The collection of the connections, the Higgs fields, and the linear maps is a point in the affine space of the small representation data.  

\subsection{Moment Map Conditions} \label{subsec:bmoment}
According to \cite{Cherkis:2010bn} the moduli space of the small bow representation is obtained by imposing the moment map conditions 
\begin{equation}\label{SmallMoment}
\mus(t,b)=\sum_{j=1}^k\Big(\delta(s-t(j))-\delta(s-h(j))\Big)\nus_j,
\end{equation}
and dividing by the action of the gauge group.  The moment map arises from considering the space of representation data, which is an affine  hyperk\"ahler space, and the natural action of the gauge group on it.  The space being hyperk\"ahler it has three symplectic structures and these are respected by the gauge transformations.    It is the three Hamiltonians $\mu_1, \mu_2,$ and $\mu_3$ generating this action that form the moment map values arranged into $\mus=\sum_a\mu_a\sigma_a$ yelding
\begin{equation}\label{SmallMap}
\mus(t,b)=-\frac{d}{ds}\ts+\sum_{j=1}^k\left(\delta(s-t(j))b_j^-\big(b_j^-\big)^\dagger+\delta(s-h(j))b_j^+\big(b_j^+\big)^\dagger\right).
\end{equation}
Within each interval this condition implies that the data satisfies the Nahm equations, which, since $t_\mu(s)$ is abelian read $\frac{d}{ds}t_i=0$ for $i=1,2,3.$ Thus within each interval $\vec{t}=(t_1,t_2,t_3)$ is constant.  At the tail $t(j)$ Eqs.~\eqref{SmallMoment} and \eqref{SmallMap} read
\begin{equation}
b_j^+ (b_j^+)^\dagger=|\vec{t}(t(j))-\vec{\nu}_j|+(\ts(t(j))-\nus_j),
\end{equation}
and at the head $h(j)$
\begin{equation}
\eeb_j \big(\eeb_j\big)^\dagger=|\vec{t}(h(j))-\vec{\nu}_j|-\Big(\ts(h(j))-\nus_j\Big).
\end{equation}
In particular these equations imply that $\ts(t(j))=\ts(h(j))$ and thus $\vec{t}(s)=\vec{t}$ is not only constant within each interval, but, has the same value  across all intervals for all values of $s.$ Once this is established let us simplify our notation slightly by introducing 
\begin{align}
\vec{t}_j&=\vec{t}-\vec{\nu}_j& &\text{and, accordingly,}&
\ts_j&=\ts-\nus_j.
\end{align}
The remaining gauge freedom can be used to completely gauge away the connection component $t_0$ within each interval, absorbing it into the phase factors of $b^\pm_j.$ At this point the calculation reduces to that of \cite{Gibbons:1996nt}.

As a result we obtain the moduli space of this small representation at level $\nus$ that is four-real-dimensional.    This space can be parameterized by $\vec{t}$ and the invariant combination of $t_0$ and complex phases of $b_j,$ leading to the Gibbons-Hawking form of the metric
\begin{equation}
ds^2=V d\vec{t}\: ^{2}+\frac{1}{V}(d\theta+\omega)^2,
\end{equation}
with $V=l+\sum_j\frac{1}{2|\vec{t_j}|},\, \theta\sim\theta+2\pi,$ and the one-form $\omega$ satisfying $*dV=-d\omega.$  Here $l$ is the sum of the lengths $l_j$ of the intervals $I_j.$  

One can now see the significance of the values $\vec{\nu}_j$ of the moment map -- these become the positions of the Taub-NUT centers.  The perimeter $2\pi/\sqrt{l}$ of the Taub-NUT circle at infinity on the other hand is determined by  the total sum of lengths of all intervals in the bow $l.$ 

Since this four-dimensional space is obtained as a moduli space of a bow representation it comes equipped with a family of self-dual connections parameterized by the union of all intervals of the bow.  In our case all of these connections are abelian instantons on $TN_k.$  These abelian instantons are instrumental in our construction and we derive them now.

\subsection{Natural Line Bundles and Self-dual Connections}
The exact abelian instanton connection will depend on how we parameterize the intervals in the bow.  Let us call the point at which $s=0$ the {\em distinguished point.} We shall be interested in the connection associated to some point $s=s_0.$  Let us call this point the {\em marked point.}

Let us consider a general position of the distinguished point on the $k$th interval, dividing it into left and right intervals on lengths $u$ and $l_0-u.$ The marked point $s_0$ is in a general position belonging to the interval number ${\rm int}(s_0):$ $s_0\in I_{{\rm int}(s_0)}.$ The distinguished point and the marked point divide the $TN_k$ bow into two parts.  Let us call the part forming the path from the distinguished point to the marked point the {\em left} path, and the part forming the path from the marked point to the distinguished point the {\em right} path. The total length of the intervals belonging to the left path is $s_0$ and the total length of the intervals belonging to the right path is $l-s_0,$ with $l=l_1+\ldots + l_k.$ 
We shall use the corresponding subscripts $l$ and $r$ to denote the quantities relating to these two parts.  For example, we denote the data of the left path by ${\rm Dat}_l$ and the data of the right path by ${\rm Dat}_r.$

The data of the bow can be viewed as the direct product of the data of the left and right paths with zero-level hyperk\"ahler reduction by the action of the gauge group $G_{s_0}$ at the marked point.  Since the moment map for $G_{s_0}$ is $\ts(s_0+)-\ts(s_0-)$ this ensures continuity at $s_0.$  Thus we have  ${\rm Dat}=({\rm Dat}_l\times{\rm Dat}_r)/\!\!/\!\!/ G_{s_0}.$  Moreover, if ${\cal G}_{s_0}$ is the group of gauge transformations that act trivially at the marked and at the distinguished point then it can be viewed as a direct product of similar groups ${\cal G}_l$ and ${\cal G}_r$ acting on the left and right path data respectively with trivial action at the marked and distinguished points.

The moduli space ${\cal M}$ of the small bow can thus be represented as a hyperk\"ahler quotient in a number of ways: 
\begin{equation}
{\cal M}={\rm Dat}/\!\!/\!\!/{\cal G}={\rm Dat}/\!\!/\!\!/({\cal G}_{s_0}\times G_{s_0})=\Big(({\rm Dat}_l/\!\!/\!\!/{\cal G}_l)\times({\rm Dat}_r/\!\!/\!\!/{\cal G}_r)\Big)/\!\!/\!\!/G_{s_0}.
\end{equation}
Here $/\!\!/\!\!/$ denotes the hyperk\"ahler reduction of \cite{Hitchin:1986ea}.
Let us denote the moduli space of respectively the left and the right paths by ${\cal M}_l$ and ${\cal M}_r$ so that ${\cal M}_l={\rm Dat}_l/\!\!/\!\!/{\cal G}_l$ and ${\cal M}_r={\rm Dat}_r/\!\!/\!\!/{\cal G}_r.$  Performing hyperk\"ahler reduction within each interval reduces the Nahm data on each interval to ${\mathbb R}^3\times S^1.$ The remaining quotient by the gauge groups acting at the ends of the intervals amounts to the quotient considered in \cite{Gibbons:1996nt} which results in a multi-Taub-NUT space.  Thus ${\cal M}_l={\rm TN}_{s_0}$ and ${\cal M}_r={\rm TN}_{l-s_0}$ with metrics
\begin{align}
ds^2_l&=V_l d\vec{t}_j^{\: 2}+\frac{1}{V_l}(d\beta+\omega_l)^2,&
ds^2_r&=V_r d\vec{t'}_j^{ 2}+\frac{1}{V_r}(d\alpha+\omega_r)^2,
\end{align}
here $\alpha$ and $\beta$ have period $2\pi$ and
\begin{align}
V_l&=s_0+\sum_{j=1}^{{\rm int}(s_0)}\frac{1}{2t_j},& V_r&=l-s_0+\sum_{j={\rm int}(s_0)+1}^k\frac{1}{2t_k},\\
&*_3d\omega_{l}= -dV_l,&
&*_3d\omega_r=- dV_r.
\end{align}
The action of the $G_{s_0}=U(1)$ is by $(\alpha, \beta)\rightarrow(\alpha-\phi, \beta+\phi)$, the invariant of this action is $\theta=\alpha+\beta$ and the moment map is $\vec{t'}_{{\rm int}(s_0)}-\vec{t}_{{\rm int}(s_0)}.$  Putting the moment map to zero we obtain the metric on the five-real-dimensional zero level set of $G_{s_0}$
\begin{equation}
ds^2=Vd\vec{t}^2+\frac{1}{V}(d\theta+\omega)^2+\frac{V}{V_l V_r}\left(d\beta+\omega_l-\frac{V_l}{V}(d\theta+\omega)\right)^2,
\end{equation}
where $V=V_l+V_r$ is the harmonic function of the $k$-centered Taub-NUT, $\omega=\omega_l+\omega_r,$  $\vec{t}=\vec{t}_{{\rm int}(s_0)}=\vec{t'}_{{\rm int}(s_0)}.$ Viewing this as a metric on the principal $U(1)_{s_0}$ bundle over ${\cal M}$ we have the natural connection $a_{s_0}$ on this bundle
\begin{equation}\label{Cona}
a_{s_0}=\omega_l-{V_l}\frac{(d\theta+\omega)}{V}.
\end{equation}
It is natural to associate the one-form connection $a^{(j)}=\omega_j-\frac{1}{2t_j}\frac{d\theta+\omega}{V},$ with $d\omega_j=-*_3 d\frac{1}{2t_j},$ to each of the Taub-NUT centers, then the above connection \eqref{Cona} in the chosen trivialization has the form
\begin{equation}\label{Connection}
a_{s}=-{s}\frac{d\theta+\omega}{V}+\sum_{j=1}^{{\rm int}(s)} a^{(j)},
\end{equation}
for $s=s_0.$  This abelian connection has self-dual curvature.  Thus each point of a bow has an associated abelian instanton given by Eq.~\eqref{Connection}.

\section{Cheshire Representation and the Monopole}\label{Sec:Cheshire}
\begin{figure}[htbp]
\begin{center}
\includegraphics[width=0.5\textwidth]{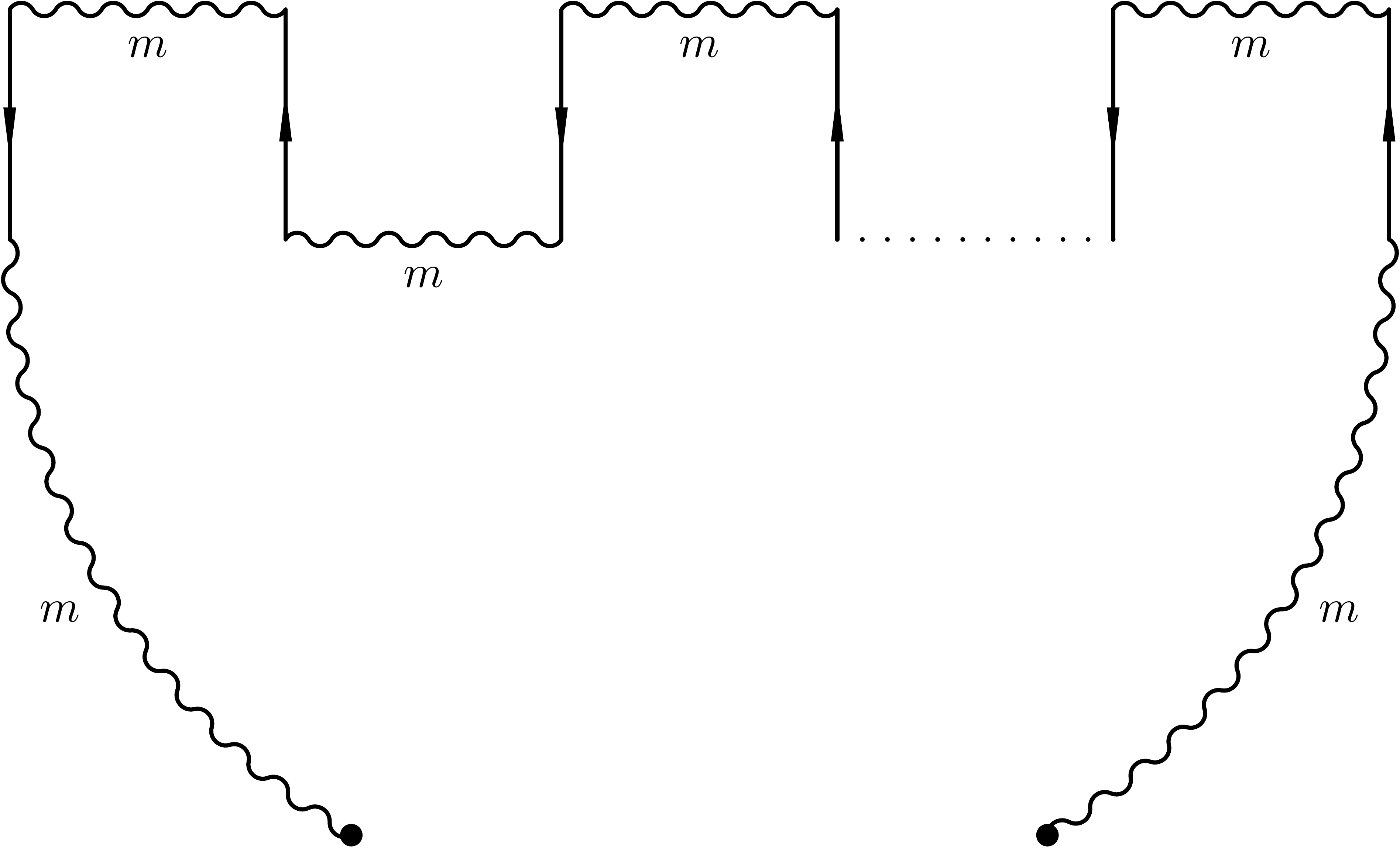}
\caption{Large bow representation: This bow has $k$ edges and $k+1$ intervals, and assigns a rank $m$ vector bundle to each of the intervals.  A solution of this bow determines a nonabelian charge $m$ monopole with $k$ Dirac singularities.}
\label{mTNLabels}
\end{center}
\end{figure}

In order to obtain a singular monopole solution of nonabelian charge $m$  we begin with the Large Representation of the $TN_k$ bow of Figure \ref{mTNLabels}. For the sake of symmetry let us choose the distinguished point with $s=0$ to be in the middle of the $k$th  interval $I_k.$  This representation has two $\lambda$-points at $s=\pm\lambda.$\footnote{This choice of $\lambda$-points makes it simpler to extract an $SU(2)$ singular monopole expression from our answer.  A priori any two points can be chosen as $\lambda$-points.} All bundles $E_j\rightarrow I_j$ have rank $m$, except the interval $I_k$ is now divided into three subintervals with the left and right subintervals each carrying  a rank $m$ bundle, while the bundle over the middle  subinterval $[-\lambda,\lambda]$ has rank zero. This latter subinterval has the $\lambda$-points as its ends.  Since the rank zero bundle has no data associated to it, this interval is not drawn in Figure \ref{mTNLabels}. This is a Cheshire representation, which ensures that the resulting instanton on the multi-Taub-NUT can be written in the form
\begin{equation}\label{Reduction}
\hat{A}=A-\Phi\frac{d\theta+\omega}{V},
\end{equation}
with $A$ and $\Phi$ independent of the variable $\theta.$  The fact that $\hat{A}$ has self-dual curvature in orientation $(dt_1,dt_2,dt_3,d\theta)$ is equivalent \cite{Kronheimer} to $A$ and $\Phi$ satisfying the Bogomolny equation $*_3F=- [D_A, \Phi].$ One can see from the form of Eq.~\eqref{Reduction} that in such a reduction of a smooth self-dual connection to a monopole the resulting monopole can have $\frac{1}{t_j}$ type singularities at the positions of the Taub-NUT centers.

More generally, if one is to construct a monopole with the $U(n)$ gauge group, one should consider a Cheshire bow with $n$ $\lambda$-points with various bundle ranks equal to the nonabelian monopole charges and, of course, one of the bundles of rank zero.  

\begin{figure}
\begin{center}
\includegraphics[width=0.5\textwidth]{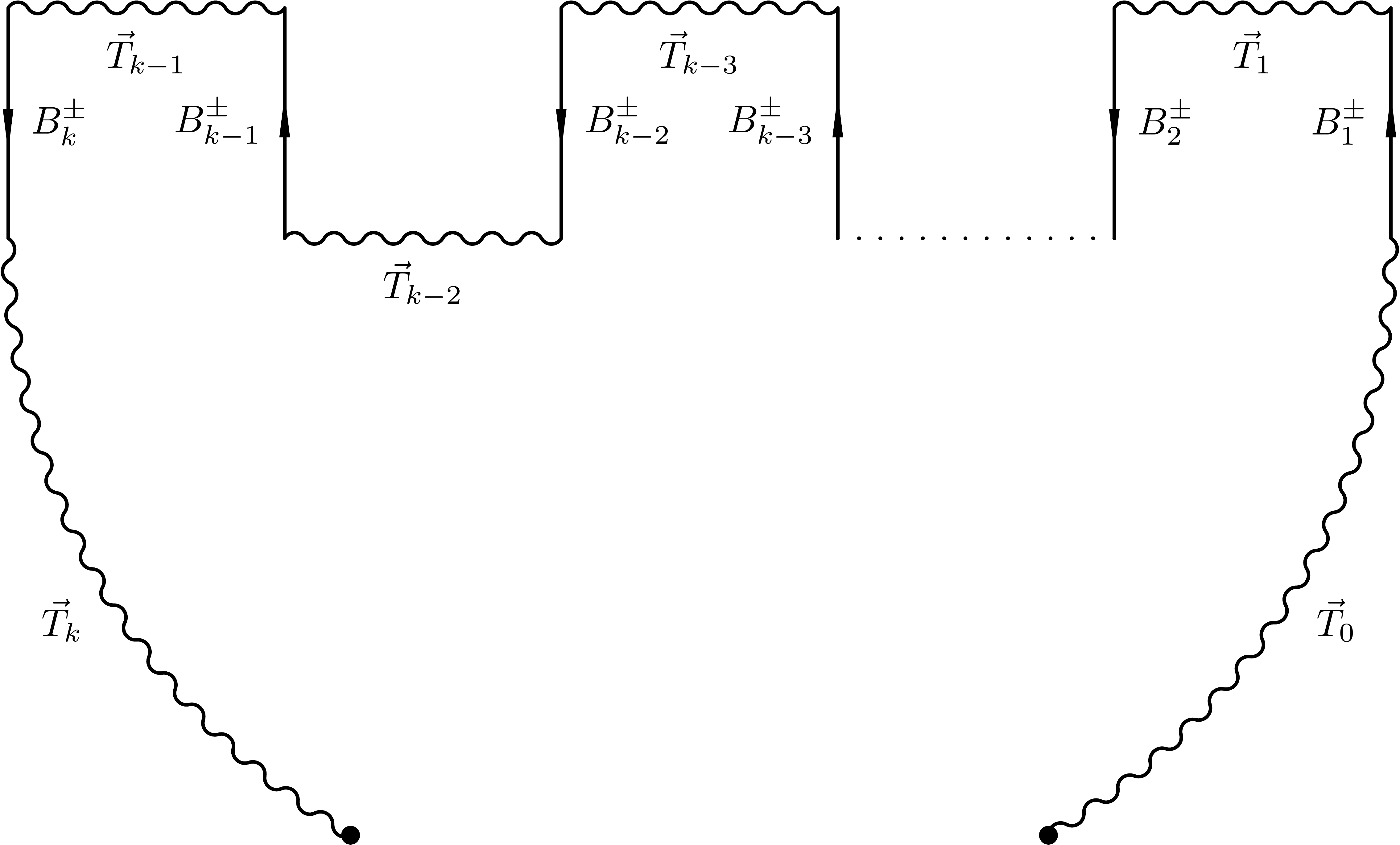}
\caption{The large bow representation with its data.  Black dots are the $\lambda$-points at $s=\pm\lambda.$}
\label{CheshireSimple}
\end{center}
\end{figure}

The data we associate to the {\em large representation} is denoted by capital letters $T$ and $B,$ as in Figure \ref{CheshireSimple}. 
As before we assign the Nahm matrix-values functions $T_1(s), T_2(s),$ and $T_3(s)$ to each interval and
to each edge we associate linear maps
\begin{equation}
B_j^{LR} : E_{t(j)} \rightarrow E_{h(j)} \qquad
B_j^{RL} : E_{h(j)} \rightarrow E_{t(j)} 
\end{equation}
which we assemble into
\begin{equation}
B_j^+  =\begin{pmatrix} \big(B_j^{RL}\big)^\dagger \\ B_j^{LR} \end{pmatrix} \qquad \eeB_j = \begin{pmatrix} \big(B_j^{LR}\big)^\dagger  \\ - B_j^{RL} \end{pmatrix}.
\end{equation}
The moment map conditions we impose for this data are 
\begin{equation}\label{BigMoment}
\mus(B,T)=-\sum_{j}\big(\delta(s-t(j))-\delta(s-h(j))\big)\nus_j,
\end{equation}
which are negative of those for the small bow of Eq.~\eqref{SmallMoment}.  Since the gauge group action on the large representation data $(T,B)$ has the same form as on the small representation data the moment map is given by the same expression, which for an arbitrary rank bow data takes the form
\begin{multline}\label{poiu}
\mus(T,B)=-\frac{d}{ds}\Ts+{\rm vec}\,\Ts\Ts\\
+\sum_{j=1}^k\left(\delta(s-t(j))B_j^-\big(B_j^-\big)^\dagger+\delta(s-h(j))B_j^+\big(B_j^+\big)^\dagger\right).
\end{multline}
Here ${\rm vec}\,\Ts\Ts=i \epsilon_{abc}[T_a, T_b]\sigma_c$, which, we note, vanishes for the rank one large representation.  At the $\lambda$-points $T_a(s)$ has to satisfy the condition  $T_a(s)=\frac{\rho(\sigma_a)}{2(s\pm\lambda)}+O((s\pm\lambda)^0,$ with $\rho$ an irreducible representation of $su(2)$ and $\sigma_a$ a Pauli matrix. 
The gauge equivalence classes of solutions to the moment map equation \eqref{BigMoment} are in one-to-one correspondence with the $U(2)$ singular monopoles with $k$ minimal singularities, with the positions of the singularities fixed to be $\vec{\nu}_j.$

For a single $U(2)$ monopole with $k$ singularities we choose the large representation with line bundles over the intervals as in 
Figure \ref{mTNLabels} putting $m=1.$  This figure for $m=1$ also illustrates the reason why our method has an advantage over the conventional Nahm transform.  In the conventional Nahm data for a monopole with $k$ singularities one has to work with the rank $k$ Nahm data, which makes it into a highly nonlinear problem.  In the Cheshire bow formulation, even though one still constructs a monopole with $k$ singularities, only abelian rank one data appears on the intervals, which makes the whole construction relatively simple.

\subsection{The Transform}
Given any solution $({T}, B)$ of the moment map conditions \eqref{BigMoment} and \eqref{poiu}, we can construct a singular monopole solution by  considering the twisted Dirac (or Weyl) operator
\begin{equation}\label{Weyl}
\mathcal{D}^\dagger = \left(\frac{d}{ds}-\ts-\Ts\right) + \sum_j \delta(s-t(j)) (\eeb_j, \eeB_j) + \sum_j \delta(s-h(j)) (B_j^+,b_j^+).
\end{equation}
This operator acts on ${\Psi}=(\psi(s), v_j)$ with $\psi(s)$ a section of $E_j\otimes e_j\otimes S,$ where $E_j\rightarrow I_j$ is the line bundle of the large representation over the interval $I_j,$ $e_j\rightarrow I_j$ is the line bundle of the small representation over the interval $I_j,$ and $S$ is the two-dimensional chiral  spin bundle, while $v_j = \begin{pmatrix} v_j^+ \\ v_j^- \end{pmatrix}$ with $v_j^+ \in e_{h(j)} \otimes E_{t(j)}$, $v_j^- \in E_{h(j)} \otimes e_{t(j)}$. So that
\begin{align}
\mathcal{D}^\dagger\Psi=\left(\frac{d}{ds}-\ts-\Ts\right)\psi 
&+ \sum_j \delta(s-t(j)) (\eeb_j v_j^+ + \eeB_j v_j^-)\nonumber\\
& + \sum_j \delta(s-h(j)) (B_j^+ v_j^+ +b_j^+ v_j^-).
\end{align}
Note that the large Nahm data in general have rank $m$ and $\Ts$ acts on $E_j\otimes S,$ so in Eq.~\eqref{Weyl} we understand $\Ts$ to be acting on $e_j\otimes E_j\otimes S$ by $1_e\otimes\Ts$ with the identity action on the small representation bundle.  Similar comments apply to $\ts, b_j^\pm,$ and $B_j^\pm$ in Eq.~\eqref{Weyl}.  We omit these $1_e$ and $1_E$ factors here to avoid cumbersome notation and also because when we specify to a charge one $U(2)$ monopole we will only deal with abelian Nahm data, in which case the above operator makes perfect sense as it is written.

The equation $\mathcal{D}^\dagger{\Psi}=0$ amounts to 
\begin{equation}\label{Bulk}
\left( \frac{d}{ds} -\ts-\Ts \right) \psi(s) = 0,
\end{equation}
within each interval and at the the interval ends
\begin{equation}\label{Match}
\psi(t(j)) = (\eeb_j, \eeB_j) v_j, \qquad \psi(h(j)) = - (B_j^+, b_j^+) v_j.
\end{equation}

If the columns of $\bf\Psi$ form an orthonormal basis of solutions of $\mathcal{D}^\dagger{\Psi}=0,$ then  the resulting self-dual connection \cite{Cherkis:2010bn} on the multi-Taub-NUT is
\begin{equation}
\hat{A}=\left({\bf\Psi},\Big(idt_a\frac{d}{dt_a}+a_s\Big){\bf\Psi}\right).
\end{equation}
Here we use the most natural norm
\begin{equation}
\left({\Psi},{\Psi}\right)=\int \psi^\dagger(s)\psi(s)ds+\sum_{j=1}^kv_j^\dagger v_j.
\end{equation}
Together with Kronheimer's reduction \eqref{Reduction} and the expression for the abelian instanton $a_s$ of Eq.~\eqref{Connection} this leads to the monopole expression
\begin{align}\label{InducedConn}
\Phi&=\left({\bf\Psi}, \Big(s+\sum_{j=1}^{{\rm int}(s)}\frac{1}{2t_j}\Big){\bf\Psi}\right),  \\
A&=\left({\bf\Psi}, \Big( idt_a\frac{d}{dt_a} + \sum_{j=1}^{{\rm int}(s)}\omega_j\Big){\bf\Psi}\right).
\end{align}

\subsection{One Singular Monopole}
We will now demonstrate the usefulness of the construction we have described by using it to obtain a charge one $U(2)$ singular monopole with $k$ singularities located at $\vec{t}=\vec{\nu}_j$, $j=1,\dots,k$. In this case the large representation is given by Figure \ref{mTNLabels} with $m=1,$ and we have abelian Nahm data $\vec{T}$ associated to each interval. The Nahm equations imply that $\vec{T}$ is constant on each interval.  The moment map condition of Eq.~\eqref{BigMoment} reads 
\begin{equation}\label{BMom}
B_j^{\pm} B_j^{\pm \dagger} = |\vec{T}+\vec{\nu}_j| \pm (\Ts + \nus_j),
\end{equation}
which implies that $\vec{T}$ is not only constant within each interval but also has the same value across all intervals.  To simplify our notation we introduce
\begin{align}
\vec{T}_j&=\vec{T}+\vec{\nu}_j,& &\text{so that}&
\Ts_j&=\Ts+\nus_j.
\end{align}
We interpret $-\vec{T}$ as the monopole position parameter, and introduce the relative position $\vec{z} = \vec{t} + \vec{T}$. In section \ref{subsec:bmoment} we also introduced the positions relative to the singularities, $\vec{t}_j = \vec{t}- \vec{\nu}_j$, and the moment map relations for the small bow were $b_j^{\pm} b_j^{\pm \dagger} = t_j \pm \ts_j$, where  $t_j = |\vec{t}_j|$.

Before we proceed solving for $\bf\Psi$ we introduce $\mathcal{P}_j = \sqrt{2(t_j T_j - \vec{t}_j \cdot \vec{T}_j )} =\sqrt{(t_j+T_j)^2 - z^2}$ and observe the following useful relations
\begin{equation}\label{Relations}
\mathcal{P}_j=B_j^{\pm\dagger} b_j^\mp = b_j^{\pm\dagger} B_j^\mp =
B_j^+ b_j^{-\dagger} + b_j^+ B_j^{-\dagger} = B_j^- b_j^{+\dagger} + b_j^- B_j^{+\dagger},
\end{equation}
\begin{equation}
(\eeb_j, \eeB_j) (\eeb_j, \eeB_j)^\dagger= T_j + t_j - \zs, \qquad 
(B_j^+, b_j^+)(B_j^+, b_j^+)^{\dagger} = T_j + t_j + \zs, 
\end{equation}
and
\begin{align}
\mathcal{P}_j &= (\eeb_j, \eeB_j) (B_j^+, b_j^+)^{\dagger} =(B_j^+, b_j^+)^{\dagger} (\eeb_j, \eeB_j), \\
\mathcal{P}_j &= (B_j^+, b_j^+) (\eeb_j, \eeB_j)^{\dagger} = (\eeb_j, \eeB_j)^{\dagger}(B_j^+, b_j^+).
\end{align}
In a way Eqs.~\eqref{Relations} hold up to a phase factor $e^{i\phi}.$  We set this factor equal to one, which amounts to choosing a gauge in which our solution will be written.

On each interval the Weyl equation of Eq.~\eqref{Bulk} is simply $\left(\frac{d}{ds} - \zs \right)\psi(s) =0$, hence within each interval $\boldsymbol{\psi}(s)=e^{s\zs}\Pi_j$ for some $s$-independent $\Pi_j,$ while the matching conditions \eqref{Match} give
\begin{align}\label{Dirv}
\mathbf{v}_j &= \frac{(B_j^+, b_j^+)^{\dagger}}{\mathcal{P}_j} \boldsymbol{\psi}(t(j))& &\text{and}& 
\boldsymbol{\psi}(h(j)) &= - \frac{T_j+t_j+\zs}{\mathcal{P}_j} \boldsymbol{\psi}(t(j)).
\end{align}
Therefore the factors $\Pi_j$ on consecutive intervals are related by
\begin{equation} \label{eqn:ConsecutivePi}
\Pi_{j} = - \frac{T_j+t_j+\zs}{\mathcal{P}_j} \Pi_{j-1},
\end{equation}
so that the choice of $\Pi_0$ (or indeed of any one of the factors $\Pi_j$) completely determines the solution $\mathbf{\Psi}.$  As we shall need an orthonormal basis of solutions we shall fix $\Pi_0$ accordingly, choosing its value so that the normalisation factor
\begin{equation} \label{eqn:NormOriginal}
\begin{split}
N^2 =\left({\bf \Psi}, {\bf \Psi}\right)= \sum_{j=0}^k \int_{p_j^L}^{p_j^R} ds\,\Pi_j^\dagger e^{2s\zs} \Pi_j + \sum_{j=1}^k \mathbf{v}_j^\dagger \mathbf{v}_j,
\end{split}
\end{equation}
is just a scalar factor (times the identity matrix $\mathbb{I}_{2\times 2}$). 

The normalised solution in this case can be written as ${\bf\Psi}_N=\frac{1}{N}{\bf\Psi}.$  Differentiating $\left({\bf\Psi}_N,{\bf\Psi}_N\right)=1,$ one verifies that 
\begin{align}
\left({\bf\Psi}_N,\frac{d}{dt_a}{\bf\Psi}_N\right)
&=\frac{1}{2}\left(\left({\bf\Psi}_N,\frac{d}{dt_a}{\bf\Psi}_N\right)-\left(\frac{d}{dt_a}{\bf\Psi}_N,{\bf\Psi}_N\right)\right)\nonumber\\
&=\frac{1}{2N^2}\left(\left({\bf\Psi},\frac{d}{dt_a}{\bf\Psi}\right)-\left(\frac{d}{dt_a}{\bf\Psi},{\bf\Psi}\right)\right).
\end{align}
These relations allow us to work with the solution ${\bf\Psi}$ satisfying $\left({\bf\Psi},{\bf\Psi}\right)=N^2$ when we compute the Higgs field and the connection below.

From Eq.~\eqref{eqn:ConsecutivePi} we see that the factor $T_j+t_j+\zs$ plays a special role in our computation, and with this in mind we observe that
\begin{equation}
T_j+t_j\pm\zs=\mathcal{P}_je^{\pm2\alpha_j\zs},
\end{equation}
where 
\begin{equation} \label{eqn:alphaj}
\alpha_j = \frac{1}{4z} \ln \frac{T_j + t_j + z}{T_j + t_j - z}.
\end{equation}
We also introduce the function $\alpha = \sum_{j=1}^k \alpha_j$ which will appear prominently in our final answer. 

We now give the expressions for the monopole fields following from Eq.~\eqref{InducedConn}. The Higgs field satisfies
\begin{equation} \label{eqn:PhiOriginal}
\begin{split}
N^2\Phi = & \sum_{j=0}^k \int_{p_j^L}^{p_j^R} ds\,\Pi_j^\dagger s e^{2s\zs} \Pi_j 
 + \sum_{j=1}^k \mathbf{v}_j^\dagger \begin{pmatrix} p_{j}^L & 0 \\ 0 & p_{j-1}^R \end{pmatrix} \mathbf{v}_j\\
&  +\sum_{j=1}^k \frac{1}{2t_j} \left( \sum_{i=j+1}^{k} \mathbf{v}_i^\dagger \mathbf{v}_i + (\mathbf{v}_j^+)^\dagger \mathbf{v}_j^+ \right) 
 + \sum_{j=1}^{k} \sum_{i=1}^{j} \frac{1}{2t_i} \int_{p_j^L}^{p_j^R} ds\,\Pi_j^\dagger e^{2s\zs} \Pi_j
\end{split}
\end{equation}
and the connection satisfies
\begin{equation} \label{eqn:AOriginal}
\begin{split}
N^2 A = & \frac{i}{2}\sum_{j=0}^k \int_{p_j^L}^{p_j^R} ds\, \left( \boldsymbol{\psi}_j^\dagger(s) d \boldsymbol{\psi}_j(s) - d\boldsymbol{\psi}_j^\dagger(s) \boldsymbol{\psi}_j(s) \right) 
 + \frac{i}{2} \sum_{j=1}^k \left( \mathbf{v}_j^\dagger d \mathbf{v}_j - d\mathbf{v}_j^\dagger \mathbf{v}_j \right)\\
& +\sum_{j=1}^k  \omega_j \left( \sum_{i=j+1}^{k} \mathbf{v}_i^\dagger \mathbf{v}_i + (\mathbf{v}_j^+)^\dagger \mathbf{v}_j^+ \right) 
 + \sum_{j=1}^{k} \sum_{i=1}^{j} \omega_i \int_{p_j^L}^{p_j^R} ds\,\Pi_j^\dagger e^{2s\zs} \Pi_j
\end{split}
\end{equation}
where $\boldsymbol{\psi}_j(s) = e^{s\zs} \Pi_j$,
with 
$\Pi_j = (-1)^j e^{2(\alpha_1 + \dots + \alpha_j)\zs} \Pi_0$
and
\begin{equation}
\mathbf{v}_j = \begin{pmatrix} \mathbf{v}_j^+ \\ \mathbf{v}_j^- \end{pmatrix} =  (-1)^j \frac{(\eeb_j, \eeB_j)^{\dagger}}{\mathcal{P}_j} e^{p_{j-1}^R \zs} e^{2(\alpha_1 + \dots + \alpha_j)\zs} \Pi_0.
\end{equation}

\begin{figure}[htbp]
\begin{center}
\includegraphics[width=0.5\textwidth]{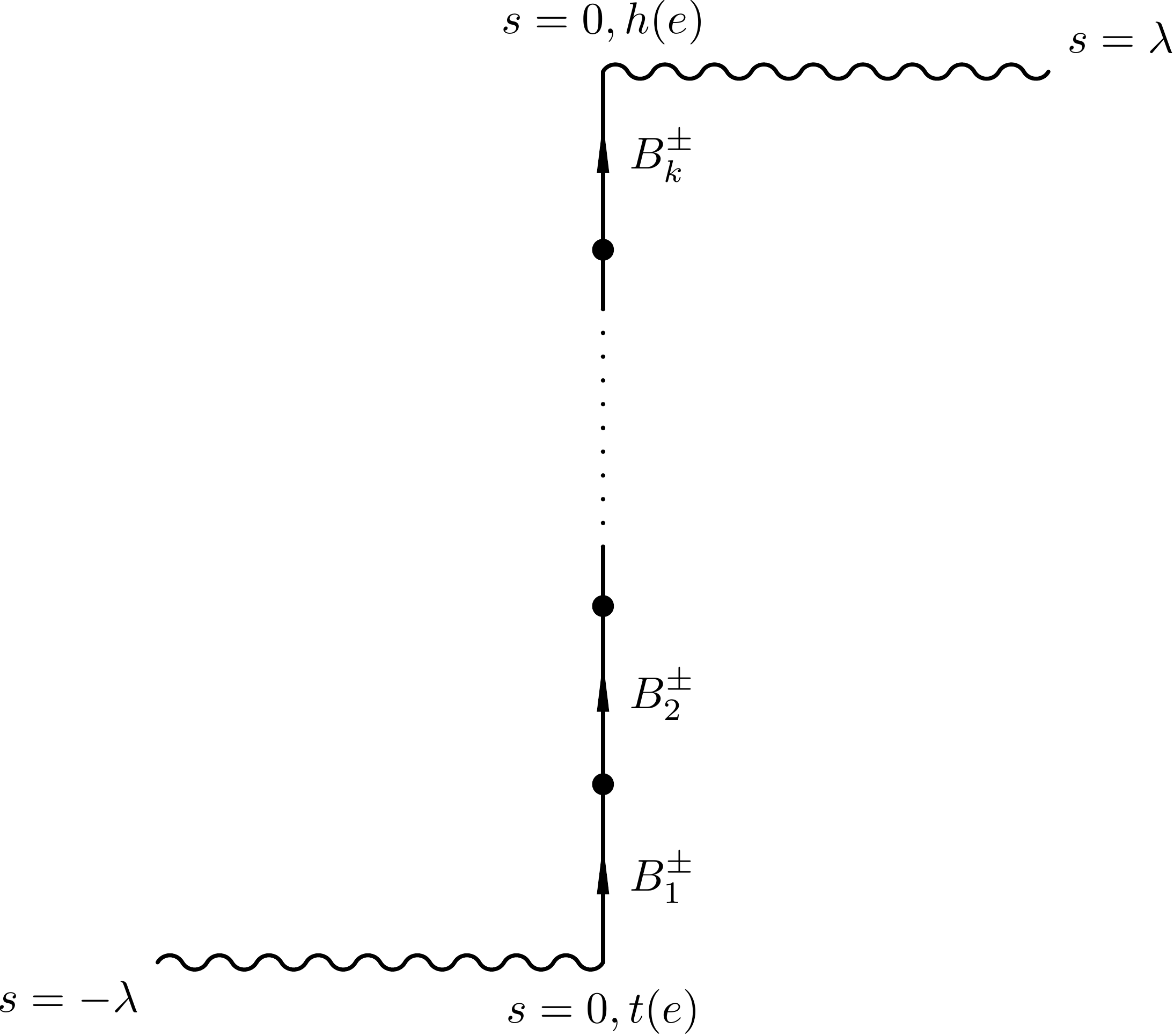}
\caption{The Cheshire representation of Figure \ref{CheshireSimple} with all but one interval shrunk to zero size.  It is important to keep in mind the relation of this diagram with the $TN_k$ bow, which is better illustrated by Figure \ref{CheshireSimple}.}
\label{SimpleCheshire}
\end{center}
\end{figure}

Except for the total length of all the intervals in the bow, the sizes of the individual intervals did not play any role in our discussion so far.  Nor will they.  From this point on we put all of the intervals in the bow to zero size with the exception of the one interval $I_k$ that contains the two $\lambda$-points.  This interval is of length $l.$ The other intervals, now shrunk to a point, are located at $s=0.$  This amounts to putting $p_0^L=-\lambda, p_k^R=\lambda,$ and all other $p_{j-1}^L=p_j^R=0,$ which substantially simplifies our computation.\footnote{Note that in contrast to the choice of the distinguished point at the beginning of this section, here we choose the distinguished point with $s=0$ to be the diametrically opposite to the middle of the interval $I_k.$} The resulting Cheshire bow representation is shown in Figure \ref{SimpleCheshire}. 

\subsubsection{Normalization}
The normalisation factor (\ref{eqn:NormOriginal}) is now given by
\begin{equation}  \label{eqn:Norm}
N^2 =(\mathbf{\Psi},\mathbf{\Psi})  = \int_{-\lambda}^0 ds\,\Pi_0^\dagger e^{2s\zs} \Pi_0 + \int_0^\lambda ds\,\Pi_k^\dagger e^{2s\zs} \Pi_k + \sum_{j=1}^k \mathbf{v}_j^\dagger \mathbf{v}_j
\end{equation}
The integrals over $s$ are straightforward and one can show that the contribution from the $s=0$ endpoints cancels with the sum of $\mathbf{v}_j$ terms; this latter calculation in fact
implies the useful relation
\begin{equation} \label{vsum}
\sum_{i=j+1}^{k} \mathbf{v}_i^\dagger \mathbf{v}_i = \frac{1}{2z^2} \left( \Pi_k^\dagger \zs \Pi_k - \Pi_{j}^\dagger \zs \Pi_{j} \right) 
\end{equation}
Using the fact that $\Pi_k = (-1)^ke^{2\alpha\zs}\Pi_0$ one ends up with an expression for $N^2$ which is proportional to $\Pi_0^\dagger e^{2\alpha \zs} \Pi_0$. This suggests a natural choice of orthogonal basis of solutions given by $\Pi_0 = e^{-\alpha \zs}.$ In this basis the normalization factor is indeed a scalar
\begin{equation}
N = \sqrt{\frac{\sinh 2 (\lambda + \alpha) z}{z} },
\end{equation}
and all basis elements have the same norm $N$ and are orthogonal to each other.

\subsubsection{Higgs Field}

The Higgs field $\Phi$ of Eq.~\eqref{eqn:PhiOriginal} becomes 
\begin{equation} \label{eqn:Phi}
\begin{split}
N^2 \Phi = &  \int_{-\lambda}^0 ds\,\Pi_0^\dagger \,s e^{2s\zs} \Pi_0  + \int_0^\lambda ds\,\Pi_k^\dagger \,s e^{2s\zs} \Pi_k \\
& + \left(\sum_{j=1}^k \frac{1}{2t_j} \right) \int_0^\lambda ds\, \Pi_k^\dagger \, e^{2s\zs} \Pi_k \\
& + \sum_{j=1}^k \frac{1}{2t_j} \left(  (\mathbf{v}_j^+)^\dagger \mathbf{v}_j^+ +\sum_{i=j+1}^{k} \mathbf{v}_i^\dagger \mathbf{v}_i \right),
\end{split}
\end{equation}
Our choice $\Pi_0=e^{-\alpha \zs}$ makes computation of the integrals especially simple, as one is now dealing only with exponentials of $2(s\pm\alpha)\zs$.
The result of the integration is
\begin{multline} 
\Phi =  \sum_{j=1}^k \frac{1}{4t_j} + \left( \left[ \lambda + \sum_{j=1}^k \frac{1}{4t_j}  \right] \coth 2(\lambda + \alpha) z -\frac{1}{2z} \right) \frac{\zs}{z} \\ 
 + \frac{z}{\sinh 2(\lambda + \alpha)z} \left[ \frac{\zs}{2z^3}  \sinh 2 \alpha z - \frac{\zs}{z^2} e^{2\alpha \zs} \sum_{j=1}^k \frac{1}{4t_j}\right.\\
\left.  +  \sum_{j=1}^k \frac{1}{2t_j} \left(   (\mathbf{v}_j^+)^\dagger \mathbf{v}_j^+ +\sum_{i=j+1}^{k} \mathbf{v}_i^\dagger \mathbf{v}_i\right) \right].
\end{multline}

The $\sum_i \mathbf{v}_i^\dagger \mathbf{v}_i$ term can be replaced with a much simpler expression using Eq.~\eqref{vsum}. After substituting $\mathbf{v}_j^+ = (-1)^j (b_j^-)^\dagger e^{-\alpha \zs}
e^{2(\alpha_1+\dots+\alpha_j)\zs}$ and bringing the remaining pieces together one finds, after some manipulation of sums of exponentials of slashed terms, that the final expression is
\begin{multline} \label{1MonHiggs}
\Phi =  \sum_{j=1}^k \frac{1}{4t_j} + \left( \bigg( \lambda + \sum_{j=1}^k \frac{1}{4t_j}  \bigg) \coth 2(\lambda + \alpha) z -\frac{1}{2z} \right) \frac{\zs}{z} \\
+ \frac{z}{ \sinh 2 (\lambda + \alpha ) z }\sum_{j=1}^k\frac{1}{2t_j \mathcal{P}_j^2} \Ts_{j\,\perp}. 
\end{multline}

The second term in the first line of this expression is reminiscent of the Higgs field of the 't Hooft-Polyakov monopole:
\begin{equation}
\Phi\big(\vec{z} \big)=\left(\lambda\coth2\lambda z-\frac{1}{2z}\right)\frac{\zs}{z}.
\end{equation}
One can see for this example that the size of the nonabelian monopole is modulated by the presence of the singularities with $ \lambda + \sum_{j=1}^k \alpha_j $ playing the role of the size controlling $\lambda$ in the 't Hooft-Polyakov case.  This size dependence and the singularity screening effect was explored in detail in \cite{Cherkis:2007jm}.

\subsubsection{Vector Potential}
From Eq.~\eqref{eqn:AOriginal} we see that the vector potential is now given by
\begin{equation} \label{eqn:A}
\begin{split}
N^2A = & \frac{i}{2} \int_{-\lambda}^0 ds\,\left( \boldsymbol{\psi}_0^\dagger(s) d \boldsymbol{\psi}_0(s) - \, h.c. \right) 
+ \frac{i}{2} \int_0^\lambda ds\,\left( \boldsymbol{\psi}_k^\dagger(s) d \boldsymbol{\psi}_k(s) - \, h.c. \right) \\
& + \sum_{j=1}^k \omega_j \int_0^\lambda ds \,\Pi_k^\dagger e^{2s\zs} \Pi_k  + \sum_{j=1}^k \omega_j \left(  (\mathbf{v}_j^+)^\dagger \mathbf{v}_j^+ +\sum_{i=j+1}^{k} \mathbf{v}_i^\dagger \mathbf{v}_i \right)\\
& + \frac{i}{2} \sum_{j=1}^k ( \mathbf{v}_j^\dagger d \mathbf{v}_j - \, h.c. ). 
\end{split}
\end{equation}
The integrals in the first line are straightforward to compute after writing $\boldsymbol{\psi}_0(s) = e^{(s-\alpha)\zs}$ and $\boldsymbol{\psi}_k(s) = (-1)^ke^{(s+\alpha)\zs}$, while the integral and the summation in the second line are the same as those that occur in the calculation of $\Phi$. One also needs 
\begin{multline} \label{eqn:vdv}
\mathbf{v}_j^\dagger d \mathbf{v}_j - d\mathbf{v}_j^\dagger \mathbf{v}_j  = \frac{1}{\mathcal{P}_j^2} e^{-( \alpha - 2[\alpha_1 + \dots + \alpha_j]) \zs} \left( b_j^- d b_j^{-\dagger} - db_j^- b_j^{-\dagger} \right) e^{- (\alpha - 2[\alpha_1 + \dots + \alpha_j]) \zs} \\
+ \frac{1}{\mathcal{P}_j} \frac{[\zs,d\zs]}{z^2} \sinh ( \alpha - 2[\alpha_1+\dots +\alpha_{j-1}] ) z\, \sinh (\alpha- 2[\alpha_1
+\dots+\alpha_j]) z,
\end{multline}
and 
\begin{equation} \label{eqn:bdb}
 b_j^- d b_j^{-\dagger} - db_j^- b_j^{-\dagger} = 2 i \omega_j (t_j - \ts_j) + \frac{1}{2t_j} [ \ts_j, d\ts_j].
\end{equation}
Using these and the explicit expression $\omega_j =  -\frac{1}{\mathcal{P}_j^2 t_j} \vec{z} \cdot ( \vec{t}_j \times d \vec{t}_j)$ it is straightforward to simplify the remaining terms obtaining the final form of the connection
\begin{equation}\label{1MonCon}
\begin{split}
A & = \frac{i}{2z} [\zs, d\zs]\left(-\frac{1}{\sinh 2(\lambda+\alpha)z} \left[ \lambda + \sum_{j=1}^k  \frac{T_j+t_j}{2\mathcal{P}_j^2}\right]+ \frac{1}{2z}\right) \\
& + \sum_{j=1}^k \frac{\omega_j}{2}  + \sum_{j=1}^k \frac{\omega_j}{2}  \frac{\zs}{z} \coth 2(\lambda +\alpha)z   + \frac{z}{\sinh 2 (\lambda + \alpha) z} \sum_{j=1}^k \frac{i  [\ts_j, d\ts_j]_\perp}{4\mathcal{P}_j^2 t_j}.
\end{split}
\end{equation}

Our results, Eqs.~\eqref{1MonHiggs} and \eqref{1MonCon}, deliver a one monopole with $k$ minimal Dirac singularities at $\vec{\nu}_j$ points.  The monopole position is parameterized by $-\vec{T}$, and we used $\vec{T}_j=\vec{T}+\vec{\nu}_j, \vec{t}_j=\vec{t}-\vec{\nu}_j,$ and $\mathcal{P}_j^2=(T_j+t_j)^2-z^2.$

\section{Conclusions}
We formulate an alternative Nahm transform for monopoles.
This new version of the Nahm transform that we apply here amounts to finding a solution $(T,B)$ of the moment maps of a large Cheshire bow representation and forming a family of Dirac operators $D^\dagger$ determined by the solution $(T,B)$ and twisted by the small representation data $(t,b).$  The moment map values of the two representations were carefully chosen to be the negatives of each other.  An orthonormal basis of solutions ${\bf\Psi}$ of the Dirac equation $D^\dagger{\Psi}=0$ gives a singular monopole with 
\begin{align}
\Phi&=\left({\bf\Psi},\bigg(s+\sum_{j\leq{\rm int} (s)}\frac{1}{2t_j}\bigg){\bf\Psi}\right),&
A&=i \left({\bf\Psi}, \nabla_a{\bf\Psi}\right) dt_a,
\end{align}
with the covariant derivative $\nabla_a=\frac{\partial}{\partial t_a}-i a_a.$
One can think of these expressions as an induced Higgs field and connection on the kernel of $D^\dagger$ from the simple abelian monopole family
\begin{align}
\phi&=s+\sum_{j=1}^{{\rm int}(s)}\frac{1}{2t_j},&
a&=\sum_{j=1}^{{\rm int}(s)}\omega_j.
\end{align}

This construction in principle delivers all singular monopoles of any charge, singularity number, and with unitary gauge group.  As an illustration, we worked out the example of one $U(2)$ monopole with $k$ singularities is complete detail.  The resulting Higgs field and connection are given in Eqs.~\eqref{1MonHiggs} and  \eqref{1MonCon}.

In \cite{Cherkis:2010kz} we use this solution to obtain an $SU(2)$ monopole with $k$ minimal singularities and analyze its properties.

\section{Acknowledgements}
The work of CB is supported in part by the undergraduate summer internship of the School of Mathematics, Trinity College Dublin.  SCh is grateful to Edward Witten for discussions.

\section{Appendix}

We describe here in detail the calculations which lead to our expressions for $\Phi$ and $A$. We have set $p_0^L=-\lambda$, $p_k^R = \lambda$ and all other points $p_j^L =
p_j^R = 0$. Solving the Dirac equation \eqref{Bulk} and \eqref{Match} we write our data $\mathbf{\Psi} = (\boldsymbol{\psi}(s), \mathbf{v}_j)$ in the form
\begin{equation}
\boldsymbol{\psi}(s) = \begin{cases} e^{s\zs} \Pi_0 & -\lambda < s < 0 \\ e^{s\zs} \Pi_k & 0 < s < \lambda \end{cases},
\end{equation}
\begin{equation}
\mathbf{v}_j = \begin{pmatrix} \mathbf{v}_j^+ \\ \mathbf{v}_j^- \end{pmatrix} = \frac{(-1)^j}{\mathcal{P}_j} (b_j^-, B_j^-)^\dagger e^{ 2[\alpha_1 + \dots + \alpha_j] \zs} \Pi_0,
\end{equation}
with $\alpha_j$ such that $\exp(2 \alpha_j z) =  \sqrt{\frac{T_j + t_j + z}{T_j + t_j -z}}$, so that
\begin{equation} 
\cosh 2 \alpha_j z = \frac{T_j + t_j}{\mathcal{P}_j}, \quad \sinh 2 \alpha_j z = \frac{z}{\mathcal{P}_j},  \label{eqn:2alpha}
\end{equation}
\begin{equation}
\cosh 4 \alpha_j z = \frac{(T_j+t_j)^2 + z^2}{\mathcal{P}_j^2}, \quad
\sinh 4 \alpha_j z = \frac{2z(T_j+t_j)}{\mathcal{P}_j^2}. \label{eqn:4alpha}
\end{equation}
Note as well from (\ref{eqn:ConsecutivePi}) that $\Pi_k=(-1)^ke^{2\alpha \zs} \Pi_0$.

\subsection*{Normalisation}

The first step in our construction is to compute the normalisation factor $N^2 = (\mathbf{\Psi}, \mathbf{\Psi}) = \int ds \,(\boldsymbol{\psi}(s))^\dagger \boldsymbol{\psi}(s) + \sum_j (\mathbf{v}_j)^\dagger \mathbf{v}_j$.    From Eq.~(\ref{eqn:Norm}) this is given by
\begin{equation}\label{Norm1} \begin{split}
N^2  = & \int_{-\lambda}^0 ds\,\Pi_0^\dagger e^{2s\zs} \Pi_0 + \int_0^\lambda ds\,\Pi_k^\dagger e^{2s\zs} \Pi_k + \sum_{j=1}^k \mathbf{v}_j^\dagger \mathbf{v}_j
\\
 = &  \frac{1}{2z} \left( \sinh 2 \lambda z \left[  \Pi_0^\dagger  \Pi_0  +  \Pi_k^\dagger  \Pi_k  \right] + \frac{1}{z} \cosh 2
\lambda z \left[ \Pi_k^\dagger \zs \Pi_k - \Pi_0^\dagger\zs  \Pi_0  \right] \right) \\
& + \frac{1}{2z^2} \Pi_0^\dagger \zs \Pi_0 - \frac{1}{2z^2} \Pi_k^\dagger \zs \Pi_k + \sum_{j=1}^k \mathbf{v}_j^\dagger \mathbf{v}_j.
\end{split}
\end{equation}
We can write the last three terms as $\frac{1}{2z^2}\Pi_0^\dagger C(k) \Pi_0$ with
\begin{equation}
C(k) =  \zs(1 - e^{4(\alpha_1+\dots+\alpha_k)\zs})   + 2 z^2 \sum_{j=1}^k  \frac{1}{\mathcal{P}_j} e^{4(\alpha_1+\dots+\alpha_j)\zs} e^{-2 \alpha_j \zs}.
\end{equation}
Then the difference $C(k)-C(k-1)$ can be written as 
\begin{equation}
C(k)-C(k-1)=e^{4(\alpha_1+\dots+\alpha_{k-1})\zs} \left( -\zs e^{4\alpha_k \zs} + \zs + \frac{2z^2}{\mathcal{P}_k} e^{2\alpha_k \zs} \right),
\end{equation}
which vanishes, as can be checked by expanding the exponentials and using the relations (\ref{eqn:2alpha}) and (\ref{eqn:4alpha}). Thus $C(k) = C(k-1)=\ldots=C(1) =  \left( -\zs e^{4\alpha_1 \zs} + \zs + \frac{2z^2}{\mathcal{P}_1} e^{2\alpha_1 \zs} \right) = 0,$
so we have shown that the last line in Eq.~\eqref{Norm1} vanishes. Hence, using $\Pi_k = (-1)^ke^{2\alpha \zs} \Pi_0$, Eq.~\eqref{Norm1} becomes 
\begin{equation} \begin{split}
N^2 &= \frac{1}{2z} \Pi_0^\dagger \left( \sinh 2 \lambda z \left(e^{4\alpha \zs} +1 \right) + \frac{\zs}{z} \cosh 2 \lambda z \left(e^{4\alpha \zs}-1 \right) \right)
\Pi_0 \\
& = \frac{1}{z}  \left( \sinh 2 \lambda z \cosh 2 \alpha z + \cosh 2 \lambda z \sinh 2 \alpha z \right)  \Pi_0^\dagger e^{2 \alpha \zs}\Pi_0 
\end{split}
\end{equation}
This expression suggests a natural choice of orthogonal basis of solutions delivered by $\Pi_0 = e^{-\alpha \zs}.$  In this basis the normalization factor satisfies  
\begin{equation}
N^2 =\frac{1}{z}\sinh 2 (\lambda + \alpha) z.
\end{equation}

\subsection*{Higgs Field}

Our Higgs field was given by (\ref{eqn:Phi}):
\begin{multline}
N^2 \Phi =   \int_{-\lambda}^0 ds\,\Pi_0^\dagger \,s e^{2s\zs} \Pi_0  + \int_0^\lambda ds\,\Pi_k^\dagger \,s e^{2s\zs} \Pi_k 
 + \left(\sum_{j=1}^k \frac{1}{2t_j} \right) \int_0^\lambda ds\, \Pi_k^\dagger \, e^{2s\zs} \Pi_k \\
 + \sum_{j=1}^k \frac{1}{2t_j} \left(  (\mathbf{v}_j^+)^\dagger \mathbf{v}_j^+ +\sum_{i=j+1}^{k} \mathbf{v}_i^\dagger \mathbf{v}_i \right) . 
\end{multline}
The integrals are straightforward to compute upon substituting $\Pi_0 = e^{-\alpha \zs}$, $\Pi_k=e^{\alpha \zs}$. One finds
\begin{multline} 
\Phi =  \sum_{j=1}^k \frac{1}{4t_j} + \left( \left[ \lambda + \sum_{j=1}^k \frac{1}{4t_j}  \right] \coth 2(\lambda + \alpha) z -\frac{1}{2z} \right) \frac{\zs}{z} \\ 
 + \frac{z}{\sinh 2(\lambda + \alpha)z} \left( \frac{1}{2z^3} \zs \sinh 2 \alpha z - \frac{\zs}{z^2} e^{2\alpha \zs} \sum_{j=1}^k \frac{1}{4t_j} \right.\\ 
\left. +  \sum_{j=1}^k \frac{1}{2t_j} \left[  (\mathbf{v}_j^+)^\dagger \mathbf{v}_j^+ +\sum_{i=j+1}^{k} \mathbf{v}_i^\dagger \mathbf{v}_i \right] \right).
\end{multline}
Now vanishing of the last three terms in Eq.~\eqref{Norm1} implies  
\begin{equation}
\sum_{i=j+1}^{k} \mathbf{v}_i^\dagger \mathbf{v}_i = \frac{1}{2z^2} \left( \Pi_k^\dagger \zs \Pi_k - \Pi_{j}^\dagger \zs \Pi_{j} \right) =
\frac{1}{2z^2} \zs e^{2\alpha \zs}  - \frac{\zs}{2z^2} e^{-2\alpha \zs} e^{4(\alpha_1 + \dots + \alpha_j) \zs} .
\end{equation} 
The Dirac equation \eqref{Dirv} gives us $v_j$ and its first component
\begin{equation}
\mathbf{v}_j^+ = \frac{(-1)^j}{\mathcal{P}_j} (b_j^-)^\dagger e^{-\alpha \zs} e^{2(\alpha_1+ \dots + \alpha_j) \zs},
\end{equation}
and a short calculation shows that
\begin{equation}
 (\mathbf{v}_j^+)^\dagger \mathbf{v}_j^+  = \frac{1}{\mathcal{P}_j^2} \Ts_{j\,\perp} + \frac{1}{\mathcal{P}_j^2} \left(t_j - \frac{\vec{z} \cdot \vec{t}_j}{z} \frac{\zs}{z} \right)  e^{-2\alpha \zs}
 e^{4(\alpha_1 + \dots + \alpha_{j}) \zs}.
\end{equation}
Combining these two observations
\begin{multline} \label{Qwe}
\Phi =  \sum_{j=1}^k \frac{1}{4t_j} + \left( \bigg( \lambda + \sum_{j=1}^k \frac{1}{4t_j}  \bigg) \coth 2(\lambda + \alpha) z -\frac{1}{2z} \right) \frac{\zs}{z} \\
+ \frac{z}{ \sinh 2 (\lambda + \alpha ) z }\sum_{j=1}^k\frac{1}{2t_j \mathcal{P}_j^2} \Ts_{j\,\perp} 
 + \frac{z}{\sinh 2(\lambda + \alpha) z} \Bigg\{ \frac{\zs}{2z^3} \sinh 2 \alpha z\\  + \sum_{j=1}^k \frac{1}{2t_j}  e^{-2\alpha \zs}  e^{4(\alpha_1 + \dots + \alpha_{j}) \zs}
\bigg(  \frac{1}{\mathcal{P}_j^2} \bigg(t_j - \frac{\vec{z} \cdot \vec{t}_j}{z} \frac{\zs}{z} \bigg)   - \frac{\zs}{2z^2} \bigg) \Bigg\}.
\end{multline}
The last line is simplified using
\begin{equation}
\frac{1}{2t_j} \left(  \frac{1}{\mathcal{P}_j^2} \left(t_j - \frac{\vec{z} \cdot \vec{t}_j}{z} \frac{\zs}{z} \right)   - \frac{\zs}{2z^2} \right) = 
-\frac{1}{2\mathcal{P}_j} \frac{\zs}{z^2} e^{-2 \alpha_j \zs},
\end{equation}
and in fact the sum of the terms in the curly brackets in Eq.~\eqref{Qwe} vanishes if
\begin{equation} \label{SinhSum}
\sinh 2\alpha z = \sum_{j=1}^k \sinh 2\alpha_j z e^{2(\alpha_1+\dots+\alpha_{j-1}) \zs - 2 (\alpha_{j+1} +\dots+ \alpha_{k}) \zs}.
\end{equation}
This is indeed the case since
\begin{multline}
\sum_{j=1}^k\left(e^{2\alpha_j\zs}-e^{-2\alpha_j\zs}\right)\, e^{2(\alpha_1+\ldots+\alpha_{j-1}-\alpha_{j+1}-\ldots-\alpha_k)\zs}=\\
\sum_{j=1}^k
\left(e^{4(\alpha_1+\ldots+\alpha_{j-1}+\alpha_j)\zs}-e^{4(\alpha_1+\ldots+\alpha_{j-1})\zs}\right)e^{-2\alpha\zs}=\\
\left(e^{4(\alpha_1+\ldots+\alpha_k)\zs}-1\right)e^{-2\alpha\zs}=
e^{2\alpha\zs}-e^{-2\alpha\zs}.
\end{multline}

\subsection*{Vector Potential}
The connection $A$ is given by Eq.~\eqref{eqn:A} so that
\begin{multline}
N^2A =  \frac{i}{2} \int_{-\lambda}^0 ds\,\left( \boldsymbol{\psi}_0^\dagger(s) d \boldsymbol{\psi}_0(s) - \, h.c. \right) 
 + \frac{i}{2} \int_0^\lambda ds\,\left( \boldsymbol{\psi}_k^\dagger(s) d \boldsymbol{\psi}_k(s) - \, h.c. \right) \\
 + \frac{i}{2} \sum_{j=1}^k ( \mathbf{v}_j^\dagger d \mathbf{v}_j - \, h.c. ) 
 + \sum_{j=1}^k \omega_j \int_0^\lambda ds \,\Pi_k^\dagger e^{2s\zs} \Pi_k 
 + \sum_{j=1}^k  \omega_j \left(  (\mathbf{v}_j^+)^\dagger \mathbf{v}_j^+ +\sum_{i=j+1}^{k} \mathbf{v}_i^\dagger \mathbf{v}_i \right).
\end{multline}
We can now insert $\mathbf{v}_j^\dagger d \mathbf{v}_j - d\mathbf{v}_j^\dagger \mathbf{v}_j$ from Eq.~\eqref{eqn:vdv} to find
\begin{multline} \label{eqn:A1}
A  = \frac{i}{2z} [\zs, d\zs]\Bigg(\frac{1}{2z} + \frac{1}{\sinh 2(\lambda+\alpha)z} \Big[ - \lambda  -\frac{\sinh 2 \alpha z}{2z} \\
 + \sum_{j=1}^k  \frac{1}{\mathcal{P}_j} \sinh ( \alpha - 2[\alpha_1+\dots +\alpha_{j-1}] ) z \sinh (\alpha- 2[\alpha_1
 +\dots+\alpha_j]) z  \Big] \Bigg) \\ 
  + \sum_{j=1}^k \frac{1}{2} \omega_j + \sum_{j=1}^k \frac{1}{2} \omega_j \frac{\zs}{z} \coth 2(\lambda +\alpha)z + \frac{z}{\sinh 2(\lambda+\alpha)z} \sum_{j=1}^k
 \frac{1}{\mathcal{P}_j^2}
 \omega_j \Ts_{j\perp} \\
  +  \frac{z}{\sinh 2(\lambda + \alpha)z}\sum_{j=1}^k \Bigg\{
 \omega_j \bigg(  \frac{1}{\mathcal{P}_j^2} \left(t_j - \frac{\vec{z} \cdot \vec{t}_j}{z} \frac{\zs}{z}
  \right) - \frac{1}{2z^2} \zs \bigg) e^{-2\alpha \zs}  e^{4(\alpha_1 + \dots + \alpha_{j}) \zs} \\
  +  \frac{i}{2\mathcal{P}_j^2} e^{-( \alpha - 2[\alpha_1 + \dots + \alpha_j]) \zs} \left( b_j^- d b_j^{-\dagger} - db_j^- b_j^{-\dagger}  \right) e^{ -(\alpha - 2[\alpha_1 + \dots + \alpha_j]) \zs}\Bigg\}. 
\end{multline}
Simple trigonometric identities and Eq.~\eqref{eqn:2alpha} give
\begin{multline}\label{eqn:sum} 
\sum_{j=1}^k  \frac{1}{\mathcal{P}_j}  \sinh ( \alpha - 2[\alpha_1+\dots +\alpha_{j-1}] ) z \sinh (\alpha- 2[\alpha_1 +\dots+\alpha_j]) z -  \frac{\sinh 2 \alpha z}{2z} = \\
-\frac{1}{2} \sum_{j=1}^k \frac{T_j+t_j}{\mathcal{P}_j^2} + \frac{1}{2}\sum_{j=1}^k \frac{1}{\mathcal{P}_j} \cosh( 2\alpha -  4[\alpha_1+\dots +\alpha_{j-1}] - 2\alpha_j) z -  \frac{\sinh 2 \alpha z}{2z}.
\end{multline}
Now,  $\cosh( 2\alpha -  4[\alpha_1+\dots +\alpha_{j-1}] - 2\alpha_j) z = \cosh [-2(\alpha_1 + \dots + \alpha_{j-1}) z + 2(\alpha_{j+1} + \dots + \alpha_k)z]$, and the sum of hyperbolic cosines in \eqref{eqn:sum} cancels against the $\sinh 2 \alpha z$ factor due to the trace part of Eq.~\eqref{SinhSum}.

This simplifies the $[\zs,d\zs]$ terms of Eq.~\eqref{eqn:A1}. Using Eq.~\eqref{eqn:bdb} for the $b_j^-$ terms and then applying $e^{\beta \zs} \as e^{\beta \zs} = \as_\perp + \frac{\vec{a} \cdot \vec{z}}{z} \frac{\zs}{z} e^{2 \beta \zs}$, $\ts_{j\perp} = - \Ts_{j \perp}$ and $\omega_j =  -\frac{1}{\mathcal{P}_j^2 t_j} \vec{z} \cdot ( \vec{t}_j \times d \vec{t}_j)$
we obtain 
\begin{equation}
\begin{split}
A & = \frac{i}{2z} [\zs, d\zs]\left(-\frac{1}{\sinh 2(\lambda+\alpha)z} \left[ \lambda + \sum_{j=1}^k  \frac{T_j+t_j}{2\mathcal{P}_j^2}\right]+ \frac{1}{2z}\right) \\
& + \sum_{j=1}^k \frac{1}{2} \omega_j + \sum_{j=1}^k \frac{1}{2} \omega_j \frac{\zs}{z} \coth 2(\lambda +\alpha)z   + \frac{z}{\sinh 2 (\lambda + \alpha) z} \sum_{j=1}^k \frac{i}{4\mathcal{P}_j^2 t_j} [\ts_j, d\ts_j]_\perp . 
\end{split}
\end{equation}

\bibliographystyle{unstr}

\end{document}